\documentclass[reprint,superscriptaddress,amsmath,amssymb,twocolumn,pra]{revtex4-2}
\makeatletter
\renewcommand{\p@subsection}{}
\makeatother
\usepackage{amsmath}
\usepackage{amsthm}
\usepackage{mathtools}
\usepackage{amssymb}
\usepackage{graphicx}
\usepackage{bm}

\usepackage{mathrsfs}
\usepackage[colorlinks, linkcolor=blue, citecolor=blue, urlcolor=blue, breaklinks=true]{hyperref}
\usepackage{color,dsfont,multirow,booktabs}
\usepackage{braket}
\usepackage{tabularx}
\usepackage{lipsum,booktabs}
\usepackage{subfiles}
\usepackage{subcaption}
\usepackage[dvipsnames]{xcolor}
\usepackage{ragged2e}
\usepackage{caption}
\makeatletter
\DeclareCaptionJustification{flushed}{%
  \leftskip\z@skip \rightskip\z@skip \parfillskip\@flushglue}
\makeatother
\usepackage{orcidlink}
\usepackage[T1]{fontenc}
\usepackage[normalem]{ulem}

\newcommand{\Rcal}{\mathcal{R}}
\newcommand{\Rinc}{\mathcal{R}^{\mathrm{inc}}}
\newcommand{\Rcoh}{\mathcal{R}^{\mathrm{coh}}}
\newcommand{\Tr}{\mathrm{Tr}}

\newcommand{\Cpath}{\mathcal{C}}

\usepackage{comment}
\begin{document}
\date{\today}
\title{Unequal Coherent Ergotropy at Equal Interferometric Visibility and Energy in Quantum-Controlled Interferometers}
\author{Tonipe Anuradha\orcidlink{0009-0001-1552-3424}}
\email{anuradhatonipe@gmail.com}
\address{International Centre for Theory of Quantum Technologies, University of Gdańsk, ul. prof. Marii Janion 4, 80-309 Gdańsk, Poland}
\author{André H. A. Malavazi\orcidlink{0000-0002-0280-0621}}
\address{International Centre for Theory of Quantum Technologies, University of Gdańsk, ul. prof. Marii Janion 4, 80-309 Gdańsk, Poland}
\author{Borhan Ahmadi\orcidlink{0000-0002-2787-9321}}
\address{International Centre for Theory of Quantum Technologies, University of Gdańsk, ul. prof. Marii Janion 4, 80-309 Gdańsk, Poland}
\author{Lucas Pollyceno\orcidlink{0000-0002-9847-1753}}
\address{International Centre for Theory of Quantum Technologies, University of Gdańsk, ul. prof. Marii Janion 4, 80-309 Gdańsk, Poland}
\author{Pedro R. Dieguez\orcidlink{0000-0002-8286-2645}}
\email{pedrodieguez@ufpr.br}
\address{Department of Physics, Federal University of Paran\'{a}, P.O. Box 19044,  81531-980 Curitiba, Paran\'{a}, Brazil}

\begin{abstract}
The ergotropy, or unitarily extractable work, of a quantum system naturally splits into incoherent and coherent contributions, respectively set by population inversion and coherences in the energy eigenbasis.
We show that this split makes the coherent ergotropy of a Mach--Zehnder qubit sensitive to path predictability that the output interference fringes do not reveal.
We send the path qubit, prepared as a thermal quantum battery, through two quantum-controlled interferometers built from the same elements, a fixed beam splitter and a beam splitter coherently controlled to be present or absent, differing only in the order in which these two act.
The two arrangements share identical visibility, stored energy, and incoherent ergotropy at every point in the parameter space.
Yet, for work extracted from the output battery alone, placing the controlled beam splitter first never yields less coherent ergotropy, and at fixed visibility and interferometric phase the difference is governed by the path predictability that this ordering creates.
The same expression reappears as a purity deficit on the beam-splitter controller, so the work advantage comes with a Landauer bound on the cost of locally resetting the controller that is never larger at the same temperature.
We further examine the effects of decoherence on the battery and consider a relativistic realization in which a Wigner rotation correlates the particle's spin with its path. Tracing out the spin decoheres the battery and allows the two arrangements to exhibit different coherent ergotropies even in regimes where their non-relativistic counterparts coincide. These results show that interference visibility is not a complete measure of the coherence available as work.
\end{abstract}
\maketitle

\section{Introduction}
\label{sec:Introduction}

Quantum batteries (QBs), quantum systems designed to store and release energy on demand, provide an operational setting in which genuinely quantum resources can be converted into useful work~\cite{alicki2013,campaioli2024colloquium,ahmadi2026LandauZener}.
The maximum work extractable from a QB through a cyclic unitary process is quantified by its ergotropy~\cite{allahverdyan2004}.
While considerable effort has been devoted to designing and optimizing charging protocols~\cite{Barra,Hu_2022,PhysRevLett.128.140501,Li:22,ahmadi2024,PhysRevA.107.032218,PhysRevA.107.042419,PhysRevA.109.012204,Rodriguez_2024,PhysRevLett.133.243602,PhysRevApplied.23.024010,ahmadi2025reservoir,PhysRevLett.134.180401,fnv6-yqmk,ahmadi2026ChiralSqueezing} and to characterizing different QB platforms~\cite{PhysRevA.106.042601,Cruz_2022,PhysRevLett.131.260401,PhysRevA.107.023725}, an equally fundamental question is how the stored work is internally organized.
For finite-dimensional systems, ergotropy admits a natural decomposition into an incoherent contribution associated with population inversion in the energy eigenbasis and a coherent contribution rooted in quantum coherence~\cite{francica2020,andre_prx_2026,wang2026ergotropy}.
This distinction is operational, as coherent ergotropy has been experimentally accessed~\cite{niu2024,li2025experimentalergotropy}, while the two components can respond differently to environmental noise~\cite{medina2025,andre_prx_2025}.
Related studies have further shown that coherent quantum control can enhance charging and work extraction through superposed trajectories, coherently controlled processes, and indefinite causal order~\cite{lai2024superposed,simonov2022switch,zhu2023ico,Simonov_2025,li2025icoergotropy}.
These developments highlight an important distinction between how much energy a battery stores and how much of that energy is actually available as useful work.

The coherent contribution makes the energy eigenbasis coherence itself a thermodynamic resource, which raises the question of how such coherence is witnessed in practice.
In interferometry, this role has traditionally been played by wave--particle duality (WPD), one of the most fundamental manifestations of quantum complementarity.
In a two-path Mach--Zehnder interferometer, wavelike behavior is characterized by the fringe visibility~$\mathcal V$, while particle-like behavior is quantified by the path predictability~$P$ or the which-way distinguishability~$\mathcal D$, which obey complementarity relations such as $P^{2}+\mathcal V^{2}\leq1$ and $\mathcal D^{2}+\mathcal V^{2}\leq1$~\cite{greenberger1988,englert_prl_1996,durr2001}.
Visibility, however, is not necessarily a complete measure of the coherence present in a state: the two coincide in particular interferometric settings, but become operationally distinct more generally~\cite{bera2015coherence,bagan2016coherence,biswas2017visibility,chrysosthemos2023visibility}.
Quantum-controlled delayed-choice interferometers offer a particularly clean setting in which to expose this distinction.
In the \emph{quantum delayed-choice experiment} (QDCE)~\cite{ionicioiu_QDCE_2011,peruzzo2012qdce}, the first beam splitter is fixed while the second is coherently controlled to be present or absent, whereas in the \emph{quantum-controlled reality experiment} (QCRE)~\cite{pedro_QCRE_2022} these roles are interchanged.
The two arrangements can produce exactly the same output visibility, yet their internal wave--particle structure differs: with the beam-splitter controller acting as the which-way detector, the QDCE carries neither path predictability nor detector distinguishability, whereas the QCRE retains genuine which-way information and supports a coexistence of partial wave and particle character~\cite{pedro_QCRE_2022,pedro_basso_QCC_2024}.
Equal interference contrast therefore does not imply equal underlying quantum resources.

In addition, the energetic side of wave--particle complementarity has been connected to thermodynamic work and information~\cite{angelo2015information}, to quantum-work statistics~\cite{xu2020dualitywork}, and to the operation of quantum heat engines~\cite{janovitch2023heatengine}.
Most directly, a single particle in a delayed-choice interferometer was proposed as a quantum battery and experimentally demonstrated an energetic relation between its wave and particle characteristics~\cite{yang2025energycoherence}.
That relation, however, links energy to the wave--particle attributes that the interferometer displays; it does not ask whether energetics can distinguish setups that display the same features.
Here we investigate exactly this scenario: two interferometers that are indistinguishable at the level of their interference fringes.
If they have the same visibility and store the same average energy, can they nevertheless differ in the work that can actually be extracted from their output states? And, if so, can this difference be traced to which-way information that is inaccessible to the interferometric visibility?

We address these questions by treating the interferometric path qubit simultaneously as a quantum battery and as a probe of wave--particle duality.
At every point of the parameter space, the QDCE and QCRE store exactly the same energy and possess the same incoherent ergotropy at the output, yet they generally differ in their coherent extractable work.
In particular, the QCRE exhibits coherent ergotropy at least as large as that of the QDCE, with its advantage arising from the additional path predictability retained by the QCRE.
The margin takes the form of a coexistence term, i.e., it is nonzero only when partial wave and particle character are simultaneously present, and its size is modulated by the interferometric phase. Since tracing out the controller turns each controlled beam splitter into a beam splitter inserted at random, this battery-side margin is fixed by the order in which the (possibly classical) mixing and the phase shift act, and does not by itself require a quantum controller; the coherence of the controller enters instead through its reset cost.
Therefore, ergotropy reveals a thermodynamic distinction between the two interferometers that visibility alone cannot detect.
We further show that this distinction has a complementary thermodynamic signature in the beam-splitter controller, which acts as the which-way detector in the QCRE.
To be reused in another cycle, the controller must be reset to its initial state, and Landauer's principle bounds the minimum work for this reset, the Landauer erasure cost, by the controller's von Neumann entropy~\cite{Georgescu2021,Chattopadhyay_2025}.
The same closed form that governs the coherent-ergotropy gap controls the difference between the controller entropies of the two protocols, so that the QCRE, which never extracts less coherent work, also never has a higher erasure cost than the QDCE.

Finally, we revisit a relativistic spin--path coupling~\cite{pedro_basso_QCC_2024}, in which the Wigner rotation of the particle's spin progressively degrades the interferometric coherence of both setups.
As the spin renders the two paths fully distinguishable, the QDCE loses all of its extractable work, whereas the QCRE still delivers a finite amount, generated at its fixed second beam splitter from the path predictability, which the spin cannot erase.
This decoherence channel thereby separates the two protocols even where their non-relativistic ergotropies coincide.
These results identify a priori path information as an operational thermodynamic resource and single out the QCRE as the more robust way of providing extractable work at the output, even when the two interferometers display identical visibility.
The remainder of this article is structured as follows. In Sec.~\ref{sec:prelim}, we introduce the model and the relevant figures of merit. In particular, we build each circuit step by step, evaluating $P$, $\mathcal{V}$, and $\mathcal{D}$ at the stage of the interferometer to which each quantifier operationally belongs. The complementarity inequalities satisfied by these quantities are also collected. In Sec.~\ref{sec:results}, the ergotropy results for both setups are presented.
In Sec.~\ref{sec:controller-cost}, we then show that the same closed form governs the state of the beam-splitter controller, and quantify the thermodynamic cost of resetting it. Then, in Sec.~\ref{sec:rel}, we extend the analysis to a relativistic regime in which a Wigner rotation correlates the path with the particle's spin. Finally, Sec.~\ref{sec:conclusion} summarizes our results and discusses future directions.

\section{Preliminaries}
\label{sec:prelim}
\begin{figure}[t]
  \includegraphics[width=\columnwidth]{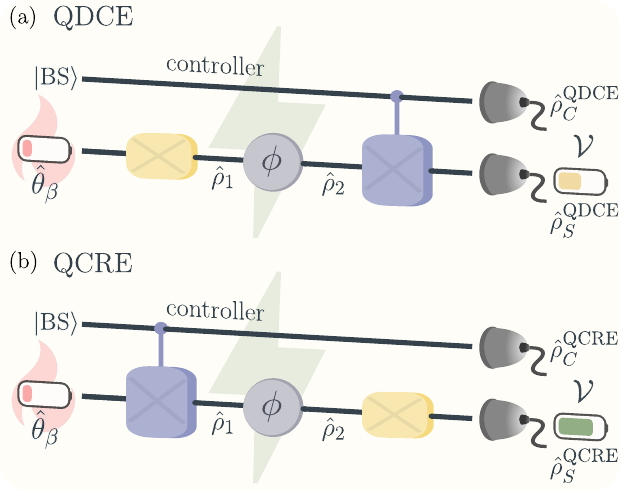}
  \caption{Schematic circuit representation of the two Mach–Zehnder interferometric setups considered. In both cases, the interferometric path qubit is initialized in the Gibbs state.
  (a)~In the QDCE, the first beam splitter (BS) is a fixed Hadamard gate (yellow box), while the second (purple box) is controlled by the BS controller.
  (b)~In the QCRE, the roles are swapped, so that the first BS is
  controlled, while the second is a fixed Hadamard. In both cases, the phase shifter (gray box) imprints a relative phase~$\phi$ on arm~$\ket{1}$.}
  \label{fig:setup}
\end{figure}
We consider a two-level system (TLS) traversing a Mach--Zehnder interferometer (MZI) whose beam-splitter (BS) configuration is governed by an auxiliary quantum degree of freedom (controller). Throughout, the TLS plays a dual role. It is the quantum battery whose extractable work is quantified by the ergotropy, and also the which-path probe whose wave--particle character is quantified by the predictability~$P$, the interferometric visibility~$\mathcal{V}$, and the which-way distinguishability~$\mathcal{D}$. Two interferometric arrangements are investigated, the quantum delayed-choice experiment (QDCE)~\cite{ionicioiu_QDCE_2011} and the quantum-controlled reality experiment (QCRE)~\cite{pedro_QCRE_2022}, represented in Fig.~\ref{fig:setup}(a) and (b), respectively.
In this paper, we set the Planck constant to $\hbar=1$.

The TLS free Hamiltonian is given by $\hat{H} = \omega\ket{1}\!\bra{1}$, where $\ket{0}$ ($\ket{1}$) is the ground (excited) state with energy~$0$ ($\omega$).
The pair $\{\ket{0},\ket{1}\}$ simultaneously labels the two spatial modes (arms) of the MZI, with $\ket{0}$ ($\ket{1}$) denoting the particle propagating in arm~$0$ (arm~$1$). 
The nondegenerate Hamiltonian assigns an energy difference $\omega$ to the two path states, allowing the interferometric degree of freedom to function as an effective quantum battery.
In the circuits of Fig.~\ref{fig:setup}, these modes are the computational basis of the path-qubit wire, so that beam splitters act as Hadamard gates and the phase shifter acts on $\ket{1}$.
The battery is initialized in the Gibbs state relative to $\hat{H}$ at inverse temperature~$\beta$,
\begin{equation}
    \hat\theta_\beta = \frac{1+q}{2}\ket{0}\!\bra{0} + \frac{1-q}{2}\ket{1}\!\bra{1},
    \label{eq:gibbs}
\end{equation}
where $q\equiv\tanh(\beta\omega/2)\in[0,1]$ interpolates between the maximally mixed state ($q=0$, $\beta\to0$) and the pure ground state $\ket{0}$ ($q=1$, reached as $\beta\to\infty$). Additionally, an auxiliary BS controller is prepared in the superposition
$\ket{\mathrm{BS}}
    = \cos\alpha\,\ket{\mathrm{in}}
    + \sin\alpha\,\ket{\mathrm{out}}$
with $\alpha\in\bigl[0,\tfrac{\pi}{2}\bigr]$ encoding whether a beam splitter is present ($\ket{\mathrm{in}}$) or absent ($\ket{\mathrm{out}}$) at the relevant port of the MZI. Hence, the full initial state of the battery--controller composite system reads
\begin{equation}
    \hat\Sigma_0
    = \hat\theta_\beta \otimes \ket{\mathrm{BS}}\!\bra{\mathrm{BS}}.
    \label{eq:rho0}
\end{equation}

Throughout, $\hat\Sigma$ denotes composite battery--controller states, whereas $\hat\rho$ denotes reduced states, with $\hat\rho_1$, $\hat\rho_2$ (inside the interferometer) and $\hat\rho_S$ (at the output) referring to the battery and $\hat\rho_C$ to the controller.
\subsection{Figures of merit}
\label{sec:figures}
All quantities below are evaluated on a general single-qubit state of
the form~\cite{andre_prx_2025} $\hat{\rho}_{S}=p_0|0\rangle\langle0|+p_1|1\rangle\langle1|+\frac{\mathcal{C}}{2}\left(e^{i\varphi/2}|0\rangle\langle1|+h.c.\right)$,
where $p_j=\bra{j}\hat \rho_S\ket{j}$ is the population of state~$\ket{j}$ and $\mathcal{C}=2|\bra{0}\hat \rho_S\ket{1}|$, with $\mathcal{C}^2\in[0,\,4p_1(1-p_1)]$, is the $l_1$-norm of coherence in the energy eigenbasis~\cite{streltsov2017}.
The ergotropy of any given state, defined with respect to a fixed Hamiltonian, is the maximum energy extractable by a cyclic unitary operation~\cite{allahverdyan2004} (see Appendix~\ref{app:ergotropy-review}).
Given $\hat{\rho}_{S}$ and $\hat{H}$, the ergotropy can be split into $\Rcal(p_1,\mathcal{C}) = \Rinc(p_1) + \Rcoh(p_1,\mathcal{C})$~\cite{francica2020,Simonov_2025, andre_prx_2025}, with an incoherent contribution due to population inversion,
\begin{equation}
    \Rinc(p_1)
    = \omega(2p_1-1)\bigl[1-\Theta(1/2-p_1)\bigr],
    \label{eq:Rinc}
\end{equation}
and a coherent component from off-diagonal elements,
\begin{align}
    \Rcoh(p_1,\mathcal{C})
    &= \frac{\omega}{2}\!\left(\psi
      -\sqrt{\psi^2-\mathcal{C}^{\,2}}\right),
    \label{eq:Rcoh}
\end{align}
where $\Theta(\cdot)$ is the Heaviside step function and $\mu[\hat \rho_S]=\mathrm{Tr}(\hat \rho_S^{2})$ is the purity, so that $\psi:= \sqrt{2\mu[\hat \rho_S]-1}$ is the length of the Bloch vector.
Hence, by construction, the incoherent part is nonzero iff $p_1>1/2$, while the coherent contribution vanishes for incoherent states ($\mathcal{C}=0$).
The three WPD quantifiers are each tied to a definite stage of the circuit.  The predictability~$P$ is read from the population imbalance of the reduced battery state $\hat \rho_1$ after the first beam splitter, i.e.~\cite{englert_prl_1996,greenberger1988}
\begin{equation}
    P = \bigl|\bra{0}\hat \rho_1\ket{0}-\bra{1}\hat \rho_1\ket{1}\bigr|.
    \label{eq:P-def}
\end{equation}
The visibility~$\mathcal{V}$ is the fringe contrast of the output detection probability $p_0(\phi)=\bra{0}\hat\rho_S\ket{0}$ at port~$0$, such that
\begin{equation}
    \mathcal{V}
    = \frac{p_{0}^{\max}-p_{0}^{\min}}{p_{0}^{\max}+p_{0}^{\min}}.
    \label{eq:V-def}
\end{equation}
The distinguishability~$\mathcal{D}$ measures how well the BS controller, acting as a which-way detector, identifies the path taken by the battery, read from the composite battery--controller state at the intermediate stage (after BS--1 and the phase shift but before BS--2).  Writing $w_j$ for the probability that the battery takes path~$j$ and $\hat \rho_C^{(j)}$ for the corresponding normalized controller state, the path information is the trace-norm distinguishability~\cite{englert_prl_1996}
\begin{equation}
    \mathcal{D}
    = \mathrm{Tr}_C\bigl|\,w_0\,\hat \rho_C^{(0)}-w_1\,\hat \rho_C^{(1)}\,\bigr|,
    \label{eq:D-def}
\end{equation}
where $|A|=\sqrt{A^\dagger A}$.
For a balanced interferometer ($w_0=w_1=1/2$) with pure detector states, this reduces to Englert's symmetric form $\mathcal{D}_{\mathrm{sym}}=\sqrt{1-|\langle d_0|d_1\rangle|^2}$. The asymmetric form is needed when the path weights are unequal, as in the QCRE. For a controller uncorrelated with the path, it collapses to $\mathcal{D}=|w_0-w_1|=P$, any excess over $P$ being the which-way information genuinely stored in the detector.
In the QDCE, the intermediate state is a product and the arms are equally weighted, so both quantities vanish. The particle character is instead probed by a blocking (retro) experiment, defined in Sec.~\ref{sec:qdce-prelim} below.

\subsection{Quantum delayed-choice experiment (QDCE)}
\label{sec:qdce-prelim}
In the QDCE~\cite{ionicioiu_QDCE_2011,pedro_basso_QCC_2024}, the first beam splitter is a fixed Hadamard gate~$\hat U_H$, while the second is controlled by the BS controller.  We build the circuit in three steps, reading off each wave--particle quantifier at the stage to which it operationally belongs.
The unitary $\hat U_H\otimes \hat{\mathbb{I}}_C$ acts on the battery alone, leaving the composite state as a product $\hat \rho_1^{\mathrm{QDCE}}\otimes\ket{\mathrm{BS}}\!\bra{\mathrm{BS}}$ with $\hat{\rho}^{\mathrm{QDCE}}_{1}=\frac{1}{2}\left(|0\rangle\langle0|+|1\rangle\langle1|\right)+\frac{q}{2}\left(|0\rangle\langle1|+h.c.\right)$.
The Hadamard converts the thermal population imbalance of $\hat \theta_\beta$, Eq.~\eqref{eq:gibbs}, entirely into off-diagonal elements, equalizing the two arms
irrespective of temperature. The predictability, Eq.~\eqref{eq:P-def},
fixed by the diagonal of $\hat{\rho}^{\mathrm{QDCE}}_{1}$, thus vanishes,
\begin{equation}
    P_{\mathrm{QDCE}} = 0.
    \label{eq:P-QDCE}
\end{equation}
The unitary $\hat U_\phi = \ket{0}\!\bra{0}+e^{i\phi}\ket{1}\!\bra{1}$ imprints a relative phase $\phi\in[0,2\pi)$ on arm~$1$, such that $\hat{\rho}^{\mathrm{QDCE}}_{2}=\frac{1}{2}\left(|0\rangle\langle0|+|1\rangle\langle1|\right)+\frac{q}{2}\left(e^{-i\phi}|0\rangle\langle1|+h.c.\right)$.
The joint unitary $\hat{\mathcal{U}}_{cH} = \hat U_H\otimes\ket{\mathrm{in}}\!\bra{\mathrm{in}} + \hat{\mathbb{I}}\otimes\ket{\mathrm{out}}\!\bra{\mathrm{out}}$ applies $\hat U_H$ to the battery conditioned on the controller being in $\ket{\mathrm{in}}$. Hence, the final battery--controller state reads
$\hat{\Sigma}^{\mathrm{QDCE}}_{f}=\frac{1+q}{2}\,\ket{\Psi^{+}_{f}}\!\bra{\Psi^{+}_{f}}+\frac{1-q}{2}\,\ket{\Psi^{-}_{f}}\!\bra{\Psi^{-}_{f}},$ with the pure-component output states $\ket{\Psi^{\pm}_{f}}=\frac{\cos\alpha}{\sqrt{2}}\bigl(\ket{x_{+}}\pm e^{i\phi}\ket{x_{-}}\bigr)\ket{\mathrm{in}}+\frac{\sin\alpha}{\sqrt{2}}\bigl(\ket{0}\pm e^{i\phi}\ket{1}\bigr)\ket{\mathrm{out}}$ where $\ket{x_\pm}=(\ket{0}\pm\ket{1})/\sqrt{2}$. Tracing out the controller and the battery, respectively, yields the following reduced states
\begin{equation}\label{eq:rho-QDCE}
    \begin{split}
        \hat{\rho}^{\mathrm{QDCE}}_{S}&=\begin{pmatrix}\frac{(1+q\cos^{2}\alpha\cos\phi)}{2} & \frac{\mathcal{C_{\mathrm{QDCE}}}}{2}\\[4pt]
\frac{\mathcal{C^{*}_{\mathrm{QDCE}}}}{2} & \frac{(1-q\cos^{2}\alpha\cos\phi)}{2}
\end{pmatrix},\\\hat{\rho}^{\mathrm{QDCE}}_{C}&=\begin{pmatrix}\cos^{2}\alpha & \tfrac{q\sin\alpha\cos\alpha\cos\phi}{\sqrt{2}}\\[4pt]
\tfrac{q\sin\alpha\cos\alpha\cos\phi}{\sqrt{2}} & \sin^{2}\alpha
\end{pmatrix},
    \end{split}
\end{equation}
with the complex off-diagonal element $\mathcal{C}_{\mathrm{QDCE}}=q(\sin^2\!\alpha\cos\phi+i\cos2\alpha\sin\phi)$, whose modulus is the coherence $\mathcal{C}$.
Since the output detection probability is
$p_0(\phi)=\tfrac{1}{2}(1+q\cos^2\!\alpha\cos\phi)$, the visibility given by Eq.~\eqref{eq:V-def} reads
\begin{equation}
    \mathcal{V}_{\mathrm{QDCE}} = q\cos^2\!\alpha .
    \label{eq:V-QDCE}
\end{equation}
Notice that the thermal factor $q$ uniformly suppresses the fringe contrast.
The intermediate state $\hat\Sigma_2^{\mathrm{QDCE}}=\hat \rho_2^{\mathrm{QDCE}} \otimes\ket{\mathrm{BS}}\!\bra{\mathrm{BS}}$ is a product for all $\alpha$, $\phi$, and $q$. The controller is therefore uncorrelated with the path and stores no which-way information, so the detector distinguishability, Eq.~\eqref{eq:D-def}, vanishes identically. With no which-way detector to interrogate, the particle character of the QDCE is instead accessed by a complementary blocking (retro) protocol, in which one arm is blocked after the first beam splitter and phase shifter, the controlled second beam splitter is kept active, and the two output probabilities are compared.
When arm~$\ket{1}$ is blocked, the surviving input is $\ket{0}$, and from $\hat{\mathcal{U}}_{cH}\ket{0}\ket{\mathrm{BS}} =\cos\alpha\,(\hat U_H\ket{0})\ket{\mathrm{in}} +\sin\alpha\,\ket{0}\ket{\mathrm{out}}$, the marginals are $P_0=\tfrac{1}{2}(1+\sin^2\!\alpha)$ and $P_1=\tfrac{1}{2}(1-\sin^2\!\alpha)$, giving $\mathcal{D}^{(1)}=|P_0-P_1|=\sin^2\!\alpha$. Blocking arm~$\ket{0}$ instead leaves $\ket{1}$. Since $\hat U_H\ket{1}=\tfrac{1}{\sqrt2}(\ket{0}-\ket{1})$ differs from $\hat U_H\ket{0}$ only by a relative sign invisible to the population detectors, while the $\ket{\mathrm{out}}$ branch now leaves the battery in $\ket{1}$, the two marginals are interchanged and $\mathcal{D}^{(0)}=\sin^2\!\alpha$ as well.
Their average defines the retrodictive distinguishability
\begin{equation}
    \mathcal{D}^{\mathrm{ret}}_{\mathrm{QDCE}}
    = \sin^2\!\alpha .
    \label{eq:D-QDCE-blocking}
\end{equation}
This quantity is independent of $q$ and $\phi$, i.e., once an arm is blocked, the surviving path is a pure path eigenstate regardless of the initial thermal mixture, and the output asymmetry is generated entirely by the controlled second beam splitter.
\subsection{Quantum-controlled reality experiment (QCRE)}
\label{sec:qcre-prelim}
In the QCRE~\cite{pedro_QCRE_2022,pedro_basso_QCC_2024}, the roles of the two beam splitters are swapped, with the first controlled by the BS controller and the second is a fixed Hadamard.  As above, we read off each quantifier at the stage to which it belongs.
The joint unitary $\hat{\mathcal{U}}_{cH}^{(1)} = \hat U_H\otimes\ket{\mathrm{in}}\!\bra{\mathrm{in}} + \hat{\mathbb{I}}\otimes\ket{\mathrm{out}}\!\bra{\mathrm{out}}$ acts first, generally establishing correlations between the battery and controller. The reduced battery state is $\hat{\rho}^{\mathrm{QCRE}}_{1}=\tfrac{1}{2}(1+q\sin^2\!\alpha)\ket{0}\!\bra{0}+\tfrac{1}{2}(1-q\sin^2\!\alpha)\ket{1}\!\bra{1}+\tfrac{q}{2}(\cos^2\!\alpha\,\ket{0}\!\bra{1}+h.c.)$.
The residual population imbalance $q\sin^2\!\alpha$ arises because, with probability $\sin^2\!\alpha$, the controller selects $\ket{\mathrm{out}}$ and no beam splitter acts, leaving the battery in its thermal state, whereas the Hadamard equalizes the arms only with probability $\cos^2\!\alpha$. Hence, the predictability, Eq.~\eqref{eq:P-def}, is nonzero,
\begin{equation}
    P_{\mathrm{QCRE}} = q\sin^2\!\alpha ,
    \label{eq:P-QCRE}
\end{equation}
with the factor $\sin^2\!\alpha$ being the weight of the no-beam-splitter branch and $q$ the initial thermal bias of $\hat \theta_\beta$.
Acting with $\hat U_\phi\otimes\hat{\mathbb{I}}_C$ on the already-correlated state gives $\hat{\rho}^{\mathrm{QCRE}}_{2}=\frac{1}{2}(1+q\sin^{2}\alpha)|0\rangle\langle0|+\frac{1}{2}(1-q\sin^{2}\alpha)|1\rangle\langle1|+\frac{q}{2}\left(e^{-i\phi}\cos^{2}\alpha|0\rangle\langle1|+h.c.\right)$.
The battery and controller are already correlated at this stage, so genuine which-way information is encoded in the controller.
The final $\hat U_H\otimes\hat{\mathbb{I}}_C$ produces $\hat\Sigma_f^{\mathrm{QCRE}}
    = \frac{1+q}{2}\,
      \ket{\Xi_f^{+}}\!\bra{\Xi_f^{+}}
    + \frac{1-q}{2}\,
      \ket{\Xi_f^{-}}\!\bra{\Xi_f^{-}}$,
with $\ket{\Xi^{+}_{f}}=\frac{\cos\alpha}{\sqrt{2}}\bigl(\ket{x_{+}}+e^{i\phi}\ket{x_{-}}\bigr)\ket{\mathrm{in}}+\sin\alpha\,\ket{x_{+}}\ket{\mathrm{out}}$, and $\ket{\Xi^{-}_{f}}=\frac{\cos\alpha}{\sqrt{2}}\bigl(\ket{x_{+}}-e^{i\phi}\ket{x_{-}}\bigr)\ket{\mathrm{in}}\notag+e^{i\phi}\sin\alpha\,\ket{x_{-}}\ket{\mathrm{out}}$.
Tracing out the controller and the battery, respectively, yields the following reduced states
\begin{equation}\label{eq:rho-QCRE}
    \begin{split}
        \hat{\rho}^{\mathrm{QCRE}}_{S}&=\begin{pmatrix}\frac{(1+q\cos^{2}\alpha\cos\phi)}{2} & \frac{\mathcal{C_{\mathrm{QCRE}}}}{2}\\[4pt]
\frac{\mathcal{C^{*}_{\mathrm{QCRE}}}}{2} & \frac{(1-q\cos^{2}\alpha\cos\phi)}{2}
\end{pmatrix},\\\hat{\rho}^{\mathrm{QCRE}}_{C}&=\begin{pmatrix}\cos^{2}\alpha & \tfrac{q\sin\alpha\cos\alpha}{\sqrt{2}}\\[4pt]
\tfrac{q\sin\alpha\cos\alpha}{\sqrt{2}} & \sin^{2}\alpha
\end{pmatrix},
    \end{split}
\end{equation}
with $\mathcal{C_{\mathrm{QCRE}}}=q(\sin^{2}\alpha+i\cos^{2}\alpha\sin\phi)$.
Notice that the diagonal entries, and hence the output detection probability $p_0(\phi)=\tfrac{1}{2}(1+q\cos^2\!\alpha\cos\phi)$, coincide with those of the QDCE, Eq.~\eqref{eq:rho-QDCE}. Thus, the two setups share the same visibility,
\begin{equation}
    \mathcal{V}\equiv\mathcal{V}_{\mathrm{QCRE}}
    = \mathcal{V}_{\mathrm{QDCE}}
    = q\cos^2\!\alpha ,
    \label{eq:V-QCRE}
\end{equation}
and differ only in their off-diagonal structure, i.e., in the coherence in the energy eigenbasis.
Here the controller is genuinely correlated with the path and acts as a which-way detector, so the trace-norm distinguishability, Eq.~\eqref{eq:D-def}, is now nonzero.  The relevant information is carried by the joint battery--controller state $\hat\Sigma_2^{\mathrm{QCRE}}$, after the controlled first beam splitter and the phase shift but before the second beam splitter. Because the controlled beam splitter biases the two arms unequally, the path weights are unbalanced, i.e.,
\begin{equation}
    w_0 = \tfrac{1}{2}(1+q\sin^2\!\alpha),
    \qquad
    w_1 = \tfrac{1}{2}(1-q\sin^2\!\alpha),
    \label{eq:w-QCRE}
\end{equation}
with $w_0-w_1=q\sin^2\!\alpha=P_{\mathrm{QCRE}}$.  Since $w_0\neq w_1$ whenever $q\sin^2\!\alpha\neq0$, the interferometer is asymmetric and
the general trace-norm expression must be employed. Evaluating it (Appendix~\ref{app:D-QCRE}) yields
\begin{equation}
    \mathcal{D}_{\mathrm{QCRE}}
    = \sin\alpha\,\sqrt{\,q^2\sin^2\!\alpha+2\cos^2\!\alpha\,},
    \label{eq:D-QCRE}
\end{equation}
which reduces to $\sin\alpha\,\sqrt{1+\cos^2\!\alpha}$ in the pure ground-state limit $q\to1$ ($\beta\to\infty$).
\begin{figure*}[t]
  \centering
  \includegraphics[width=1.0\textwidth]{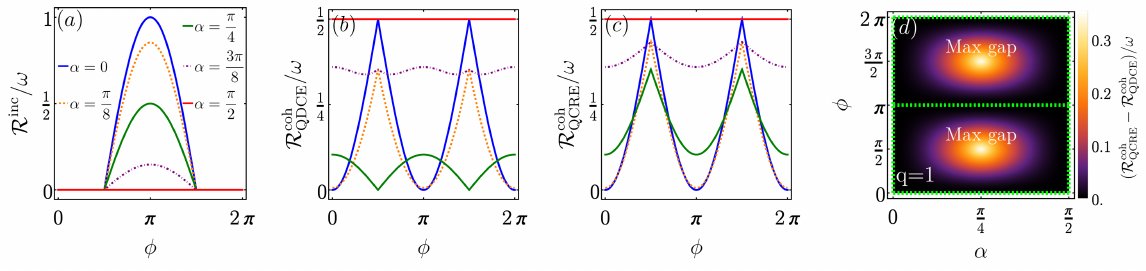}
 \caption{(a)--(c) Output ergotropy components as functions of the phase~$\phi$ for different values of the beam-splitter mixing angle~$\alpha$ (legend), and (d) coherent ergotropy gap, at $q=1$. (a)~Incoherent ergotropy $\Rinc/\omega$, common to both setups since $\Rinc_{\mathrm{QDCE}}=\Rinc_{\mathrm{QCRE}}$ [Eq.~\eqref{eq:Rinc-equal}], and nonzero only for $\cos\phi<0$, vanishing at $\alpha=\pi/2$. (b)~Coherent ergotropy of the QDCE, $\Rcoh_{\mathrm{QDCE}}/\omega$ [Eq.~\eqref{eq:Rcoh-QDCE}], which is phase-dependent through $G(\alpha,\phi)$ and dips to zero at $\alpha=\pi/4$, $\phi=\pi/2,\,3\pi/2$. (c)~Coherent ergotropy of the QCRE, $\Rcoh_{\mathrm{QCRE}}/\omega$ [Eq.~\eqref{eq:Rcoh-QCRE}].
 The QCRE coherent ergotropy is everywhere at least as large as that of the QDCE. (d) Heatmap of the coherent ergotropy gap $\Rcoh_{\mathrm{QCRE}}-\Rcoh_{\mathrm{QDCE}}$ on the $(\alpha,\phi)$ plane (units of~$\omega$, $q=1$).  The gap vanishes on the boundary lines (green dashed) where $\sin\phi=0$ or $\sin(2\alpha)=0$, in agreement with Eq.~\eqref{eq:gap-identity}, and peaks at $\alpha=\pi/4$, $\phi=\pi/2,\,3\pi/2$ (white dots), reaching $\omega/(2\sqrt{2})\approx0.354\,\omega$.
}
  \label{Fig2}
\end{figure*}
\subsection{Complementarity bounds}
\label{sec:bounds}
The quantifiers obtained above obey the standard complementarity inequalities of wave--particle duality.
For any interferometer equipped with a which-way detector, here the BS controller, the detector distinguishability and the visibility satisfy~\cite{englert_prl_1996,greenberger1988}
\begin{equation}
    \mathcal{V}^2 + \mathcal{D}^2 \leq 1 .
    \label{eq:EGY}
\end{equation}
For the QDCE the controller is uncorrelated with the path, so the detector distinguishability vanishes and the bound reduces to its purely wavelike form,
\begin{equation}
    \bigl(\mathcal{V}^2+\mathcal{D}^2\bigr)_{\mathrm{QDCE}}
    = \mathcal{V}^2 = q^2\cos^4\!\alpha \leq 1 ,
    \label{eq:EGY-QDCE}
\end{equation}
saturated only in the closed-interferometer, zero-temperature limit, i.e., $\alpha\to0$, $q\to1$. For the QCRE the controller is a genuine detector, and substituting Eqs.~\eqref{eq:V-QCRE} and~\eqref{eq:D-QCRE} gives
\begin{equation}
    \bigl(\mathcal{V}^2+\mathcal{D}^2\bigr)_{\mathrm{QCRE}}
    = \frac{1+q^2}{2}
      - \frac{1-q^2}{2}\cos^2(2\alpha) .
    \label{eq:EGY-QCRE}
\end{equation}
This saturates the bound, $\mathcal{V}^2+\mathcal{D}^2=1$, only in the pure limit $q\to1$.
The predictability and visibility obey~\cite{durr2001}
\begin{equation}
    P^2 + \mathcal{V}^2 \leq 1 .
    \label{eq:Durr}
\end{equation}
With the values found above,
\begin{align}
    \bigl(P^2+\mathcal{V}^2\bigr)_{\mathrm{QDCE}}
      &= q^2\cos^4\!\alpha \leq 1 ,
    \\
    \bigl(P^2+\mathcal{V}^2\bigr)_{\mathrm{QCRE}}
      &= q^2\bigl(\sin^4\!\alpha + \cos^4\!\alpha\bigr)
       \leq 1 ,
\end{align}
the QCRE saturating the D\"urr bound only as $q\to1$ and at $\alpha=0,\pi/2$ (minimum $q^{2}/2$ at $\alpha=\pi/4$), whereas the EGY bound is saturated for every $\alpha$ as $q\to1$.  Thermal noise thus
degrades the path knowledge and the fringe contrast simultaneously.

Although the QDCE controller carries no path information, $\mathcal{D}_{\mathrm{QDCE}}=0$, it retains a particle character accessible by the blocking protocol of Sec.~\ref{sec:qdce-prelim}, quantified by
$\mathcal{D}^{\mathrm{ret}}_{\mathrm{QDCE}}=\sin^2\!\alpha$.  This retrodictive quantity satisfies an EGY-type relation which, since no detector is involved, does not follow from Englert's theorem but holds directly because $\cos^4\!\alpha+\sin^4\!\alpha\le1$,
\begin{equation}
    \mathcal{V}^2+\bigl(\mathcal{D}^{\mathrm{ret}}_{\mathrm{QDCE}}\bigr)^2
    = q^2\cos^4\!\alpha + \sin^4\!\alpha \leq 1,
    \label{eq:EGY-ret}
\end{equation}
saturated only at $\alpha=\pi/2$, or at $\alpha=0$ with $q\to1$, and whose particle term is independent of temperature, i.e., once an arm is blocked, the surviving path is a pure eigenstate, so the retrodictive distinguishability is immune to the thermal mixing that suppresses $\mathcal{V}$ and $P$.
In the detector channel, the QDCE is purely wavelike at all temperatures, $P_{\mathrm{QDCE}}=\mathcal{D}_{\mathrm{QDCE}}=0$, with the entire resource carried by $\mathcal{V}$, while its particle character is confined to the retrodictive channel of Eq.~\eqref{eq:EGY-ret}.
The QCRE, by contrast, genuinely distributes the resource between the particle-like quantities $P$, $\mathcal{D}$ and the wavelike visibility $\mathcal{V}$, with the balance set by $\alpha$. Increasing $\alpha$ enhances the $\ket{\mathrm{out}}$ branch and trades $\mathcal{V}$ for $P$, and for $\mathcal{D}$ as well at $q=1$. The blocking protocol, by contrast, carries no particle information for the QCRE. Once an arm is blocked after the controlled first beam splitter and the phase shift, the surviving path eigenstate enters the fixed Hadamard, which yields equal output probabilities, so that $\mathcal{D}^{\mathrm{ret}}_{\mathrm{QCRE}}=0$ for all $q$, $\alpha$, and $\phi$. The particle character of the two protocols thus resides in complementary channels, the retrodictive one for the QDCE and the predictive (detector) one for the QCRE, and only the latter feeds the coherent-ergotropy gap of Sec.~\ref{sec:comparison}.

\section{Results}
\label{sec:results}
To connect wave--particle duality with the system energetics, we compute the extractable work at two different stages of each interferometer.
The predictability $P$ is fixed inside the interferometer, whereas the visibility $\mathcal{V}$ is read off the output fringes, and neither coincides in general with the coherence or population imbalance that supply $\mathcal{R}^{\mathrm{coh}}$ and $\mathcal{R}^{\mathrm{inc}}$. We therefore compute the ergotropy directly from the reduced states and only then express it through $\mathcal{V}$ and $P$.
More specifically, at each stage, we apply Eqs.~\eqref{eq:Rinc}--\eqref{eq:Rcoh} and express the result through the wave--particle quantifiers $\mathcal{V}$ and $P$
previously established in Sec.~\ref{sec:prelim}.
\subsection{QDCE}
\label{sec:results-QDCE}
Here, the two paths are still open, and no information about which one was taken is available, so the battery is expected to hold its energy in its purely coherent form. This is confirmed by the explicit state, since after Step 2, the reduced battery state has
\begin{equation}
    p_{1,\mathrm{in}} = \tfrac{1}{2},
    \qquad
    \mathcal{C}_{\mathrm{in}} = q,
    \qquad
    \psi_{\mathrm{in}} = q.
    \label{eq:in-QDCE-inputs}
\end{equation}
The Hadamard has transferred the thermal population bias entirely into off-diagonal coherence, leaving a flat diagonal, so that coherence is the sole repository of extractable work at this stage.
Indeed, since $p_{1,\mathrm{in}}=\tfrac{1}{2}$ and $\psi_{\mathrm{in}}^2-\mathcal{C}_{\mathrm{in}}^2=0$, the incoherent ergotropy vanishes and the extractable work is purely coherent, i.e.\ $\Rinc_{\mathrm{in}}=0$ and
\begin{equation}
    \Rcal_{\mathrm{in}}
    = \Rcoh_{\mathrm{in}}
    = \frac{\omega}{2}\,q .
    \label{eq:R-in-QDCE}
\end{equation}
The second beam splitter recombines the two paths and, because it is applied conditionally on the control qubit, it interpolates between a closed and an open interferometer. Coherence is then partially converted back into a population imbalance, and the ergotropy acquires an incoherent component whose activation is governed by the sign of the accumulated phase. Given Eq.~\eqref{eq:rho-QDCE} after Step 3, one has
\begin{align}
    p_1 &= \tfrac{1}{2}(1-q\cos^2\!\alpha\cos\phi),
    \\[2pt]
    \mathcal{C}
    &= q\sqrt{\cos^2(2\alpha)\sin^2\!\phi+\sin^4\!\alpha\cos^2\!\phi},
    \\[2pt]
    \psi &= qG(\alpha,\phi).
\end{align}
where
\begin{equation}
    G(\alpha,\phi)
    \equiv\sqrt{(\cos^4\!\alpha+\sin^4\!\alpha)\cos^2\!\phi
        + \cos^2(2\alpha)\sin^2\!\phi}.
    \label{eq:Gdef}
\end{equation}
A direct computation gives the key identity $\sqrt{\psi^2-\mathcal{C}^2}=\mathcal{V}|\cos\phi|$.
Thus, for all $\phi$,
\begin{align}
    \Rinc &= \omega\,\mathcal{V}|\cos\phi|\,\Theta(-\cos\phi),
    \label{eq:Rinc-QDCE}\\[4pt]
    \Rcoh &= \frac{\omega}{2}
       \bigl[qG(\alpha,\phi)-\mathcal{V}|\cos\phi|\bigr].
    \label{eq:Rcoh-QDCE}
\end{align}
Both components are therefore governed by the visibility. While $\Rinc$ activates
through $\Theta(-\cos\phi)$ and is modulated by $\mathcal{V}|\cos\phi|$, $\Rcoh$
measures the excess of $qG(\alpha,\phi)$ over the same modulation.
The step function reflects the fact that only for $\cos\phi<0$ does the recombination push population into the excited level, creating the inversion that incoherent extraction requires, whereas for $\cos\phi\ge0$ no inversion occurs and the ergotropy is entirely coherent.
\begin{figure}[t]
  \centering
  \includegraphics[width=\linewidth]{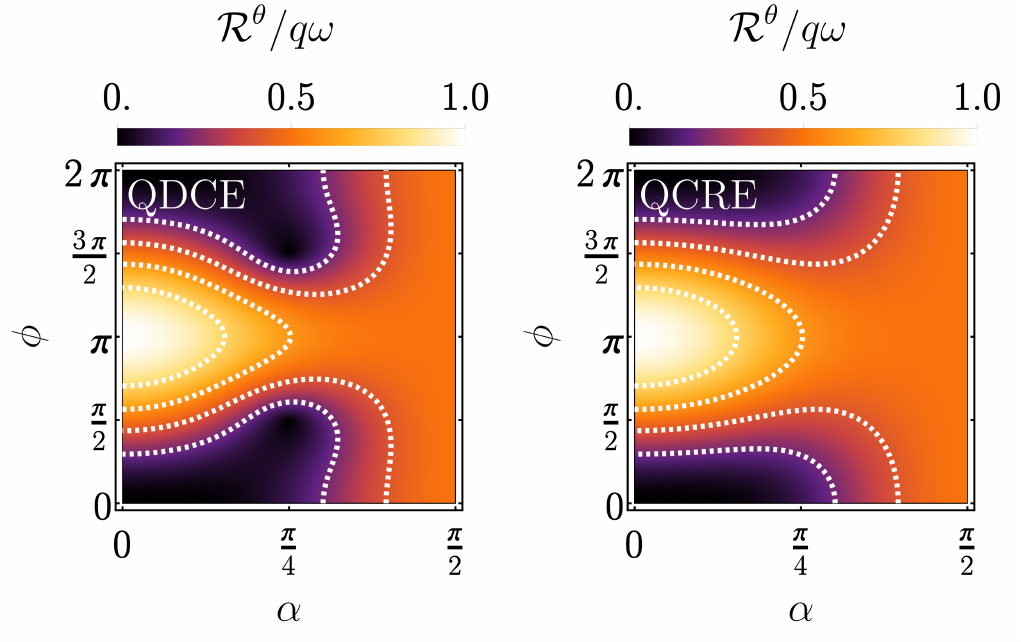}
  \caption{Contour maps of the total ergotropy $\Rcal/q\omega$ on the $(\alpha,\phi)$ plane for QDCE (left) and QCRE (right). Dotted white curves are iso-ergotropic contours~\cite{andre_prx_2026}. The thermal factor $q$ only rescales the map, so that $\Rcal^{\theta}/q$ is independent of the initial temperature [Eq.~\eqref{eq:thermal-scaling}] for $q>0$, extended to $q=0$ by continuity. Both peak at the qubit maximum for $\alpha=0$, $\phi=\pi$, and are pinned at $\omega q/2$ along $\alpha=\pi/2$. They differ in their zeros, since the QDCE output becomes passive at $\alpha=\pi/4$, $\phi=\pi/2$ and $3\pi/2$ (dark regions), where the QCRE still yields $\omega q/(2\sqrt{2})$.}
  \label{fig3}
\end{figure}
\subsection{QCRE}
\label{sec:results-QCRE}
Unlike the QDCE, the state inside the QCRE already carries which-path information, so the coherent work is reduced by the predictability rather than set by $q$ alone.
Similarly, after Step 2, one has
\begin{align}
    p_{1,\mathrm{in}} &= \tfrac{1}{2}(1-q\sin^2\!\alpha)= \tfrac{1}{2}(1-P),
    \\[2pt]
    \mathcal{C}_{\mathrm{in}} &= q\cos^2\!\alpha = \mathcal{V},
    \\[2pt]
    \psi_{\mathrm{in}}
    &= q\sqrt{\cos^4\!\alpha+\sin^4\!\alpha}
     = \sqrt{\mathcal{V}^2+P^2}.
    \label{eq:psi-in-QCRE}
\end{align}
Here $p_{1,\mathrm{in}}\le\tfrac{1}{2}$ for all $\alpha$ and $q$, and $\sqrt{\psi_{\mathrm{in}}^2-\mathcal{C}_{\mathrm{in}}^2}
=P$, which implies that $\Rinc_{\mathrm{in}}=0$ and
\begin{equation}
    \Rcoh_{\mathrm{in}}
    = \frac{\omega}{2}\bigl(\sqrt{\mathcal{V}^2+P^2}-P\bigr).
    \label{eq:Rcoh-in-QCRE}
\end{equation}
The inside coherence equals the output visibility, $\mathcal{C}_{\mathrm{in}}=\mathcal{V}$, and the passive-state gap equals the predictability~$P$. The coherent ergotropy is thus governed by how far the Pythagorean combination $\sqrt{\mathcal{V}^2+P^2}$ exceeds~$P$.
After Step 3, Eq.~\eqref{eq:rho-QCRE} gives
\begin{align}
    p_1 &= \tfrac{1}{2}(1-q\cos^2\!\alpha\cos\phi),
    \\[2pt]
    \mathcal{C}
    &= q\sqrt{\sin^4\!\alpha+\cos^4\!\alpha\sin^2\!\phi},
    \\[2pt]
    \psi
    &= q\Lambda(\alpha)
     = \sqrt{\mathcal{V}^2+P^2}.
\end{align}
where $\Lambda(\alpha)\equiv\sqrt{\cos^4\!\alpha+\sin^4\!\alpha}$. In contrast to the QDCE, the output purity of the QCRE is not modulated by the interferometric phase, i.e., $\psi$ is independent of~$\phi$. Moreover, the output population coincides with that of the QDCE. Thus, the same identity holds,
$\sqrt{\psi^2-\mathcal{C}^2}=\mathcal{V}|\cos\phi|$, giving the following ergotropy decomposition
\begin{align}
    \Rinc &= \omega\,\mathcal{V}|\cos\phi|\,\Theta(-\cos\phi),
    \label{eq:Rinc-QCRE}\\[4pt]
    \Rcoh &= \frac{\omega}{2}
       \bigl[\sqrt{\mathcal{V}^2+P^2}-\mathcal{V}|\cos\phi|\bigr].
    \label{eq:Rcoh-QCRE}
\end{align}
Since $\psi$ is fixed, the phase does not alter the purity of the output but only the extent to which coherence is converted into population imbalance, which is what shifts weight from $\Rcoh$ to $\Rinc$.
\subsection{Comparison and wave--particle mapping}
\label{sec:comparison}
In Sec.~\ref{sec:prelim}, we showed that the two circuits share the same visibility throughout the parameter space, i.e., $\mathcal{V}=q\cos^2\!\alpha$.
However, equal visibility does not necessarily imply equal coherent ergotropy, as the output states may differ in their off-diagonal elements.
The output populations coincide, so both the internal energy, $\langle \hat H\rangle=\langle \hat H\rangle_{\mathrm{QDCE}}=\langle \hat H\rangle_{\mathrm{QCRE}} =\omega p_1$, and incoherent ergotropy are identical,
\begin{equation}
    \Rinc_{\mathrm{QDCE}} = \Rinc_{\mathrm{QCRE}}
    = \omega\,\mathcal{V}|\cos\phi|\,\Theta(-\cos\phi).
    \label{eq:Rinc-equal}
\end{equation}
This component is therefore a purely wavelike quantity: it scales with~$\mathcal{V}$ and vanishes whenever $\mathcal{V}=0$, i.e.\ at $\alpha=\pi/2$  [Fig.~\ref{Fig2}(a)] or $q=0$.

\begin{figure}[t]
    \centering
    \includegraphics[width=0.48\textwidth]{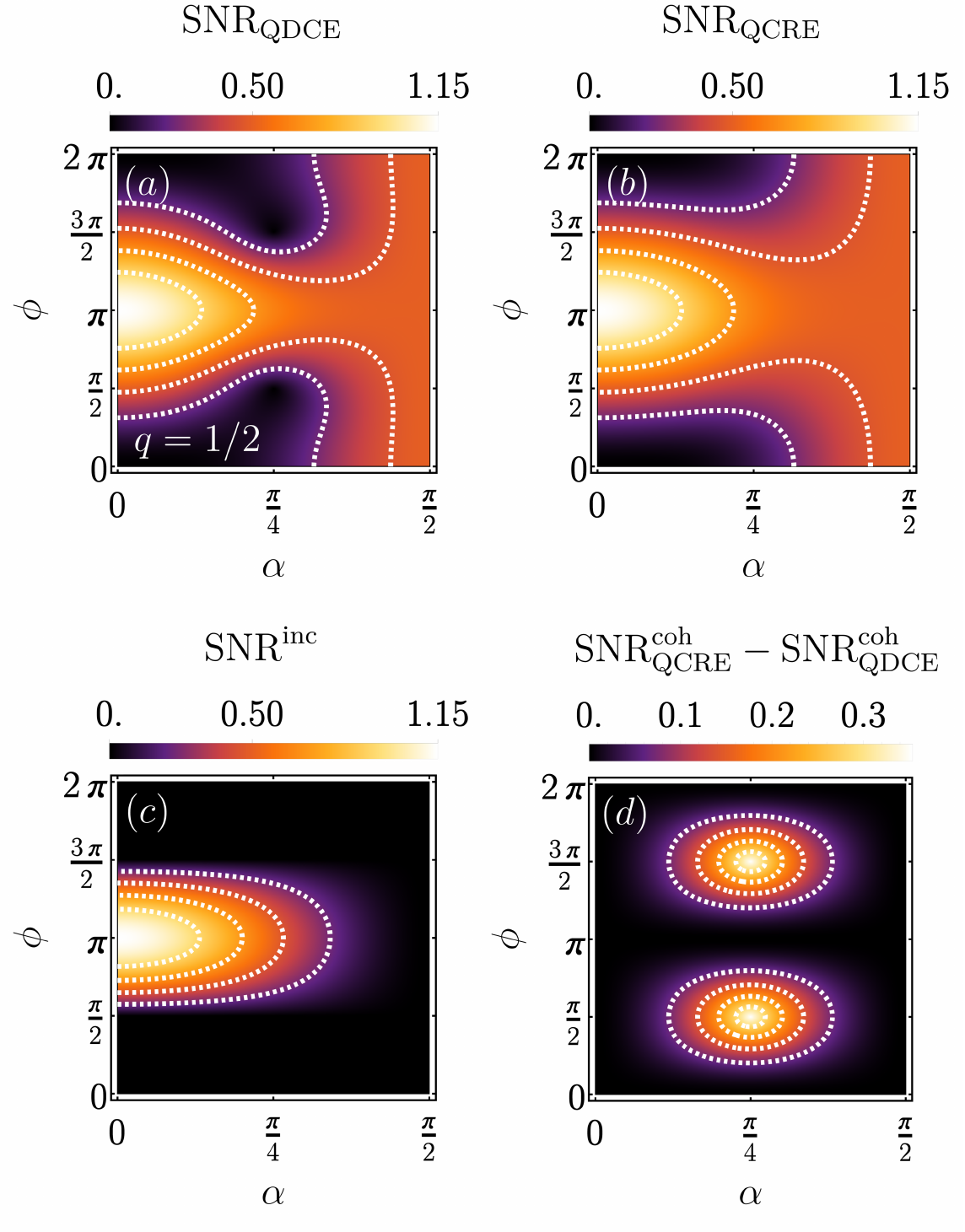}
    \caption{Work--like signal--to--noise ratio $\textrm{SNR}$ for the (a) QDCE and (b) QCRE setups, and its equal (c) incoherent component, for $q=1/2$. (d) $\textrm{SNR}^{\rm coh}$ gap.
    }
    \label{fig4}
\end{figure}
\begin{table*}[t]
\centering
\caption{Ergotropy and work-to-noise ratio in the limiting cases, with $\mathcal{V}=q\cos^{2}\!\alpha$ and $\Lambda(\alpha)=\sqrt{\cos^{4}\!\alpha+\sin^{4}\!\alpha}$. Entries are the total $\Rcal=\Rinc+\Rcoh$, and in the full-wave row the expression given is valid on both halves of the phase cycle. The last column gives the QCRE--QDCE difference in the two figures of merit. The SNR is dimensionless, with
$\sigma_H=\tfrac{\omega}{2}\sqrt{1-\mathcal{V}^{2}\cos^{2}\!\phi}$.}
\label{tab:limits}
\begin{ruledtabular}
\begin{tabular}{lccccc}
Limit
  & $\Rcal_{\mathrm{QDCE}}/\omega$
  & $\Rcal_{\mathrm{QCRE}}/\omega$
  & $\mathrm{SNR}_{\mathrm{QDCE}}$
  & $\mathrm{SNR}_{\mathrm{QCRE}}$
  & Gap $\bigl(\Rcal/\omega\,;\,\mathrm{SNR}\bigr)$ \\
\colrule
$\phi=0$
  & $\tfrac{q}{2}\bigl[\Lambda-\cos^2\!\alpha\bigr]$
  & $\tfrac{q}{2}\bigl[\Lambda-\cos^2\!\alpha\bigr]$
  & $\dfrac{q\Lambda-\mathcal{V}}{\sqrt{1-\mathcal{V}^{2}}}$
  & $\dfrac{q\Lambda-\mathcal{V}}{\sqrt{1-\mathcal{V}^{2}}}$
  & $0$ \\[10pt]
$\phi=\pi$
  & $\tfrac{q}{2}\bigl[\Lambda+\cos^2\!\alpha\bigr]$
  & $\tfrac{q}{2}\bigl[\Lambda+\cos^2\!\alpha\bigr]$
  & $\dfrac{q\Lambda+\mathcal{V}}{\sqrt{1-\mathcal{V}^{2}}}$
  & $\dfrac{q\Lambda+\mathcal{V}}{\sqrt{1-\mathcal{V}^{2}}}$
  & $0$ \\[10pt]
$\alpha=0$ (wave)
  & $\tfrac{q}{2}(1-\cos\phi)$
  & $\tfrac{q}{2}(1-\cos\phi)$
  & $\dfrac{q(1-\cos\phi)}{\sqrt{1-q^{2}\cos^{2}\!\phi}}$
  & $\dfrac{q(1-\cos\phi)}{\sqrt{1-q^{2}\cos^{2}\!\phi}}$
  & $0$ \\[10pt]
$\alpha=\pi/2$ (particle)
  & $\tfrac{q}{2}$ & $\tfrac{q}{2}$ & $q$ & $q$ & $0$ \\[2pt]
$\alpha=\pi/4,\ \phi=\pi/2,\,3\pi/2$
  & $0$ & $q/(2\sqrt{2})$ & $0$ & $q/\sqrt{2}$
  & $q/(2\sqrt{2})\,;\ q/\sqrt{2}$ \\
\end{tabular}
\end{ruledtabular}
\end{table*}
Since $\sqrt{\psi^2-\mathcal{C}^2}=\mathcal{V}|\cos\phi|$ holds for both circuits, the entire thermodynamic distinction between them is carried by the purity factor~$\psi$,
\begin{equation}
    \psi_{\mathrm{QDCE}} = qG(\alpha,\phi),
    \qquad
    \psi_{\mathrm{QCRE}} = q\Lambda(\alpha),
    \label{eq:psi-WPD}
\end{equation}
which affects their coherent ergotropy.
For the QCRE, $\psi$ is simply the Pythagorean combination of visibility and predictability within~$\Lambda(\alpha)$ and is independent of the phase $\phi$, whereas for the QDCE it is modulated by~$G(\alpha,\phi)$. The coherent sector consequently splits, with a gap given by
\begin{equation}
\begin{split}
    \Rcoh_{\mathrm{QCRE}} - \Rcoh_{\mathrm{QDCE}}
    &= \frac{\omega}{2}
      \bigl[\sqrt{\mathcal{V}^2+P^2}-qG(\alpha,\phi)\bigr]
    \\[4pt]
    &= \frac{\omega}{2}\,
      \frac{2\,\mathcal{V}P\sin^2\!\phi }
           {\sqrt{\mathcal{V}^2+P^2}+qG(\alpha,\phi)}
    \ \ge 0 .
\end{split}
    \label{eq:Rcoh-diff}
\end{equation}
Notice that the sign directly follows from the identity
\begin{equation}
\begin{split}
    \mathcal{V}^2+P^2 - q^2G^2(\alpha,\phi)
      &= \frac{q^2}{2}\sin^2(2\alpha)\sin^2\!\phi \\[3pt]
      &= 2\,\mathcal{V}P\sin^2\!\phi \ \geq 0 ,
\end{split}
    \label{eq:gap-identity}
\end{equation}
in which the two duality quantifiers enter as a product. The thermodynamic distinction between the protocols is therefore a coexistence effect governed by $\mathcal{V}P\sin^2\phi$, which vanishes in the pure-wave ($P=0$) and pure-particle ($\mathcal{V}=0$) limits and is nonzero only where both characters are simultaneously present. Explicitly, $q^2G^2=\mathcal{V}^2+P^2-2\mathcal{V}P\sin^2\!\phi$, so all ergotropies depend only on $(\mathcal{V},P,\phi)$ and, at fixed visibility and phase, the gap is set by $P$ alone. The ergotropy thus carries no information beyond these quantities; what it adds is access to $P$ at the output, where the fringes do not reveal it.
Hence, $\Rcoh_{\mathrm{QCRE}}\ge\Rcoh_{\mathrm{QDCE}}$, with equality only on the boundary lines $\phi\in\{0,\pi,2\pi\}$ and $\alpha\in\{0,\pi/2\}$, or trivially at $q=0$, and with the
maximum gap $\omega q/(2\sqrt{2})$ at $\alpha=\pi/4$, $\phi=\pi/2,3\pi/2$, where $\Rcoh_{\mathrm{QDCE}}$ vanishes altogether. This structure is displayed in Figs.~\ref{Fig2}(b)--(d). The QDCE coherent ergotropy dips to zero at the balanced, maximally shifted points, while the QCRE one does not, and the heatmap of the gap is bounded by the green dashed lines predicted by Eq.~\eqref{eq:gap-identity}, with the maximum gap at  $\phi=\pi/2$ and $3\pi/2$ (white spots).
Notice that, at fixed visibility, a nonzero predictability enables the QCRE advantage, while the interferometric phase controls how strongly that additional particle-like structure appears as coherent extractable work. The advantage is, however, not monotonic in $P$. At $\phi=\pi/2$, for instance, the gap reduces to $\tfrac{\omega}{2}\bigl(\sqrt{\mathcal{V}^2+P^2}-|\mathcal{V}-P|\bigr)$, which at fixed visibility is largest for $P=\mathcal{V}$, a value attainable only for $\mathcal{V}\le1/2$, since $\mathcal{V}+P=q\le1$. In the pure limit $q\to1$ ($\beta\to\infty$), where the QCRE saturates the Englert bound $\mathcal{V}^2+\mathcal{D}^2=1$ [Eq.~\eqref{eq:EGY-QCRE}], the QCRE purity may be written $\psi_{\mathrm{QCRE}}=\sqrt{1-\mathcal{D}^2+P^2}$, exhibiting the purity that feeds its coherent ergotropy as increasing with predictability and decreasing with distinguishability. At finite temperature, however, the exact form $\sqrt{\mathcal{V}^2+P^2}$ must be retained, since $\mathcal{V}^2+\mathcal{D}^2<1$ there.

Two remarks fix the origin and the scope of this ordering. First, tracing out the controller after the controlled Hadamard maps any battery state $\hat\rho$ onto $\cos^2\!\alpha\,\hat U_H\hat\rho\,\hat U_H+\sin^2\!\alpha\,\hat\rho$, which is exactly what a beam splitter inserted at random with probability $\cos^2\!\alpha$ produces. All battery-side quantities above ($\mathcal V$, $P$, $\Rinc$, $\Rcoh$ and their gap) are therefore unchanged if the controller is replaced by a classical random variable, in line with classical causal models of the QDCE statistics~\cite{chaves2018causal}. The gap is set by the order of this mixing relative to the phase shift, and it requires the a priori predictability $P$ rather than the which-way record of the detector: for a dephased controller, Eq.~\eqref{eq:D-def} gives $\mathcal D_{\mathrm{QCRE}}=P_{\mathrm{QCRE}}$, while the gap is unchanged. The coherence of the controller matters instead for its reset cost (Sec.~\ref{sec:controller-cost}). Second, the ordering refers to unitaries acting on the output battery alone. With access to the battery--controller pair and a degenerate controller, both final states are unitarily equivalent to $\hat\theta_\beta\otimes\ket{\mathrm{BS}}\!\bra{\mathrm{BS}}$, so the global ergotropy is $\omega p_1$ in both protocols. Inside the interferometer the ordering is reversed, $\mathcal R^{\mathrm{in}}_{\mathrm{QDCE}}=\omega q/2\ge\mathcal R^{\mathrm{in}}_{\mathrm{QCRE}}$ [Eqs.~\eqref{eq:R-in-QDCE} and~\eqref{eq:Rcoh-in-QCRE}]; it is the final beam splitter that converts the predictability of the QCRE into output coherence.

Since $\hat \theta_\beta$ [Eq.~\eqref{eq:gibbs}] is a convex combination of $\ket{0}\!\bra{0}$ and $\ket{1}\!\bra{1}$, all three inputs scale uniformly with~$q$,
\begin{equation}
p_1^{\theta}-\tfrac12 = q\bigl(p_1^{|0\rangle}-\tfrac12\bigr),\quad
\mathcal{C}^{\theta} = q\,\mathcal{C}^{|0\rangle},\quad
\psi^{\theta} = q\,\psi^{|0\rangle},
\end{equation}
where the superscript $\ket{0}$ denotes the quantity evaluated for the pure input. Therefore, the ergotropy inherits the scaling,
\begin{equation}
    \mathcal{R}^{\theta}(\alpha,\phi) = q\,\mathcal{R}^{|0\rangle}(\alpha,\phi),
    \label{eq:thermal-scaling}
\end{equation}
i.e., at both stages and for both setups, the initial temperature acts as a global suppression factor for every ergotropy, and none of the comparisons above is altered by it.
Figure~\ref{fig3} shows this behavior for the QDCE (left) and QCRE (right) setups.

To compare the extractable work with the output energy fluctuations, we introduce the work-like signal-to-noise ratio
$\mathrm{SNR}=\mathcal R/\sigma_H$, where
$\sigma_H=\sqrt{\mathrm{Var}(\hat H)}$ is the standard deviation of the battery energy~\cite{ahmadi2026ChiralSqueezing}.
Since the QDCE and QCRE have identical output populations, they also have identical energy fluctuations. Their SNR ordering therefore follows directly from their ergotropic ordering, with the incoherent contribution common to both protocols and their entire difference residing in the coherent sector.
As shown in Figs.~\ref{fig4}(a) and \ref{fig4}(b), the QCRE consequently achieves an SNR that is never smaller than that of the QDCE throughout parameter space. Figure~\ref{fig4}(c) displays their common incoherent contribution, whereas Fig.~\ref{fig4}(d) isolates the coherent advantage,
$\mathrm{SNR}^{\rm coh}_{\rm QCRE}-\mathrm{SNR}^{\rm coh}_{\rm QDCE}\ge0$.
For the representative value $q=1/2$, this advantage is maximal at
$\alpha=\pi/4$ and $\phi=\pi/2,3\pi/2$, where the QDCE contribution vanishes while the QCRE remains finite.
Thermal mixing suppresses the SNR monotonically, although the suppression is not simply linear in $q$, unlike that of the ergotropy itself, because the energy variance is also temperature dependent.

Table~\ref{tab:limits} collects the extremal points of the parameter space. We highlight the following points: First, $\Rinc$ is non-zero only where $\cos\phi<0$. It appears in the $\phi=\pi$ row, where $\Rinc=\omega\mathcal{V}$, and over half the cycle of the full-wave row, becoming the entire ergotropy at $\alpha=0$, $\phi=\pi$. In the remaining rows $\Rinc=0$, and the comparison concerns the coherent sector alone. Second, the two protocols agree wherever Eq.~\eqref{eq:gap-identity} vanishes, namely at trivial phase $\phi\in\{0,\pi\}$, in both the full-wave and full-particle limits, and in the infinite-temperature limit $q=0$.
A QCRE advantage therefore requires both a nontrivial phase and a genuine superposition of the two circuit settings. Third, the extreme cases of the duality are not the extreme cases of the extractable work. The full-wave limit reaches the qubit maximum $\omega q$ at $\phi=\pi$, the full-particle limit is pinned at $\omega q/2$ by the fixed Hadamard, and the largest disagreement between the protocols occurs in between, at the balanced setting $\alpha=\pi/4$ with $\phi=\pi/2,\,3\pi/2$.
Fourth, the same ordering holds for the work--to--noise ratio. Since $\sigma_H=\omega\sqrt{p_1(1-p_1)}$ depends only on the population, which the two protocols share, every boundary line above carries over unchanged to the SNR. At $\phi=\pi/2$ the battery is unbiased, $\sigma_H=\omega/2$, and the ratio reduces to the Bloch length itself, so that $\mathrm{SNR}_{\mathrm{QDCE}}=q|\cos(2\alpha)|$ vanishes at $\alpha=\pi/4$ while $\mathrm{SNR}_{\mathrm{QCRE}}=q\Lambda$ remains finite, giving the maximal ratio gap $q/\sqrt{2}$. In the full-particle limit both are pinned at $q$, whereas in the full-wave limit, at $\phi=\pi$, the ratio is $2q/\sqrt{1-q^{2}}$. The point of maximal ergotropy is also a point of minimal energy fluctuation, and the ratio diverges as $q\to1$, where the output becomes the pure inverted state $\ket{1}$, and the fluctuation vanishes altogether.

\section{Controller entropy and the reset cost of complementarity}
\label{sec:controller-cost}
Section~\ref{sec:comparison} showed that, at equal visibility, the coherent-ergotropy advantage of the QCRE requires the extra path predictability~$P$ it carries. This is a priori path information, encoded in the battery's unequal path weights, Eq.~\eqref{eq:w-QCRE}, rather than stored in a detector. The controlled first beam splitter that generates it, however, also correlates the BS controller with the path (Sec.~\ref{sec:qcre-prelim}), so that the controller holds path-conditioned information, quantified by $\mathcal{D}_{\mathrm{QCRE}}\geq P_{\mathrm{QCRE}}$, and acts as a physical which-way detector instead of a passive agent.
Therefore, reusing the controller to charge a further battery requires resetting it to $\ket{\mathrm{BS}}$, and Landauer's bound assigns an intrinsic thermodynamic cost for this process~\cite{Georgescu2021}.
We now examine how this cost differs between the two protocols.

Given equations~\eqref{eq:rho-QDCE} and~\eqref{eq:rho-QCRE} for the controller states, we observe that they share the same population imbalance $\cos(2\alpha)$ and differ only in their coherence. More specifically, the QCRE coherence is phase-independent, while the QDCE is modulated by $\cos\phi$. This happens since the QCRE's entangling gate is the first operation of the circuit, acting on the initial Gibbs input, whereas the QDCE's is the last, acting on a battery already reshaped by the phase shift.

Defining the controller purity length as in Eq.~\eqref{eq:Rcoh}, $\psi_C=\sqrt{2\mu[\hat\rho_C]-1}$ with $\mu[\hat\rho_C]=\mathrm{Tr}(\hat\rho_C^{2})$, both cases collapse to $\psi_C^{2}=\cos^{2}(2\alpha)+\tfrac{q^{2}}{2}\sin^{2}(2\alpha)\,c^{2}$, where $c=1$ for the QCRE and $c=\cos\phi$ for the QDCE, so that
\begin{equation}
\begin{split}
    \psi_{C,\mathrm{QCRE}}^{2}-\psi_{C,\mathrm{QDCE}}^{2}
      &= \frac{q^{2}}{2}\sin^{2}(2\alpha)\sin^{2}\!\phi \\[3pt]
      &= 2\,\mathcal{V}P\sin^{2}\!\phi ,
\end{split}
    \label{eq:controller-purity-identityMAIN}
\end{equation}
which is precisely the gap identity, Eq.~\eqref{eq:gap-identity}, now evaluated on the controller instead of the battery. For both protocols, the battery and controller purity lengths differ by the same protocol- and phase-independent offset, $\psi^2-\psi_C^2=-(1-q^2)\cos^2 2\alpha$, so Eq.~\eqref{eq:controller-purity-identityMAIN} follows directly from Eq.~\eqref{eq:gap-identity}.
The purity the QCRE battery retains, relative to the QDCE, is exactly the purity the QDCE controller loses.

 The von Neumann entropy of a qubit, $S(\hat \rho)=H\bigl(\tfrac{1+\psi}{2}\bigr)$ with $H(x)=-x\ln x-(1-x)\ln(1-x)$, is monotonically decreasing in the purity-length~$\psi$. Equation~\eqref{eq:controller-purity-identityMAIN} therefore gives
\begin{equation}
    S\bigl(\hat \rho_C^{\mathrm{QDCE}}\bigr)\ge S\bigl(\hat \rho_C^{\mathrm{QCRE}}\bigr),
    \label{eq:S-ordering}
\end{equation}
with equality exactly where the coherent-ergotropy gap itself vanishes, namely on the boundary lines $\sin\phi=0$ or $\sin(2\alpha)=0$, and at $q=0$.
If the BS controller is a reusable physical resource, with reset to $\ket{\mathrm{BS}}$ before the next charging cycle, Landauer's principle bounds the minimum work required for that reset by~\cite{Georgescu2021,Chattopadhyay_2025}
\begin{equation}
    W_{\mathrm{erase}} \ge \beta^{-1}\; S(\hat\rho_C),
    \label{eq:landauer}
\end{equation}
which we refer to as the Landauer erasure cost of the controller.
We assume the controller has trivial (degenerate) energy levels and both protocols are reset at the same inverse temperature~$\beta$ as the battery.
Here, we consider a local reset of the controller to a pure state, without exploiting correlations with the battery.
\begin{figure}[t]
    \centering
    \includegraphics[width=0.65\linewidth]{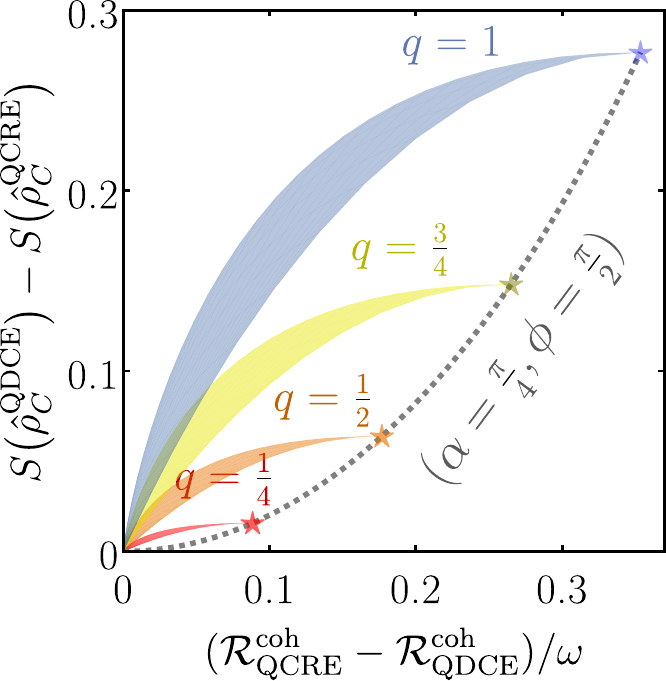}
    \caption{Controller entropy difference $S(\hat\rho_C^{\mathrm{QDCE}})-S(\hat\rho_C^{\mathrm{QCRE}})$ versus the coherent-ergotropy gap $(\Rcoh_{\mathrm{QCRE}}-\Rcoh_{\mathrm{QDCE}})/\omega$, for $\alpha\in[0,\pi/2]$ and $\phi\in[0,2\pi]$, at four values of $q$. Stars mark the extreme points $\alpha=\pi/4$, $\phi=\pi/2,\,3\pi/2$, and the dotted curve traces them as $q$ varies.}
    \label{fig5}
\end{figure}

By Eq.~\eqref{eq:S-ordering}, the Landauer bound on the QDCE controller-reset work is never smaller than that of the QCRE, and is strictly larger away from the equality conditions. The QCRE therefore delivers at least as much coherent ergotropy as the QDCE [Eq.~\eqref{eq:gap-identity}], with a work-to-noise ratio likewise never lower [Fig.~\ref{fig4}], while leaving its own which-way detector in a state that costs no more, and generally less, to recycle.
The coherent-ergotropy gap and the entropy saving grow together, as Fig.~\ref{fig5} shows. Both quantities vanish on the same parameter boundaries and reach their maxima at $\alpha=\pi/4$, $\phi=\pi/2,\,3\pi/2$.
Thus, the maximal work advantage and the maximal controller entropy difference occur at the same setting. The region is a band rather than a curve, so a given gap does not fix the entropy difference uniquely, and it contracts towards the origin as $q$ decreases, the whole trade-off being suppressed at higher temperature.
The two protocols also differ in when the controller entropy is generated. A unitary confined to one subsystem leaves the reduced state of its complement unchanged, so $\hat\rho_C$ can change only where the entangling (controlled-Hadamard) gate acts, which is Step~1 in the QCRE (Sec.~\ref{sec:qcre-prelim}) and Step~3 in the QDCE. The QCRE controller therefore reaches its final value, and its final entropy, immediately. In contrast, the QDCE controller stays exactly pure through Steps~1--2, with $\hat\rho_C^{\mathrm{QDCE}}\big|_{\mathrm{inside}}=\ket{\mathrm{BS}}\!\bra{\mathrm{BS}}$ and $S=0$, and acquires its entropy, generically larger than the QCRE's, only at the last step, where it correlates with a battery already reshaped by the phase. Inside the interferometer, the ordering of Eq.~\eqref{eq:S-ordering} therefore inverts, the QCRE controller having already acquired its final entropy, whereas the QDCE controller remains pure, a timing asymmetry that follows directly from the swapped beam-splitter order defining the QDCE/QCRE pair.
The which-way distinguishability~$\mathcal{D}$ of Sec.~\ref{sec:figures} answers a different operational question from $S(\hat \rho_C)$. $\mathcal{D}$ fixes the success probability, $P_{\mathrm{succ}}=(1+\mathcal{D})/2$, of a single optimal (Helstrom) measurement on the controller aimed at guessing the path -- an epistemic, one-shot quantity. $S(\hat \rho_C)$, by contrast, is the physical structure the controller carries regardless of whether it is ever measured, and it is what enters the Landauer bound above. The two are generically inequivalent -- $\mathcal{D}_{\mathrm{QCRE}}$, Eq.~\eqref{eq:D-QCRE}, is not a function of $\psi_{C,\mathrm{QCRE}}$ alone -- because $\mathcal{D}$ is built from the difference of the path-conditioned controller states while $\hat \rho_C$ is their sum.
Distinguishability and controller entropy are thus two inequivalent readouts of the battery--controller correlations. The former quantifies what an observer could learn about the path in a single trial, whereas the latter quantifies the cost of erasing all the correlations the controller holds with the battery, whether or not they carry path information. The distinction is sharpest in the QDCE, whose controller stores no which-way information, $\mathcal{D}_{\mathrm{QDCE}}=0$, yet ends with the larger entropy, since its correlations are created by the final controlled beam splitter, at the output, rather than between the arms.

\section{A concrete decoherence channel: relativistic spin--path entanglement}
\label{sec:rel}
The results so far assumed a battery that traverses the interferometer with its path coherence affected only by the initial thermal mixedness, encoded in~$q$. However, any physical realization is prone to coherence loss through coupling to unmonitored degrees of freedom.
This effect can be taken into account by considering a decoherence channel independent of the thermal preparation.
We now show that the battery-side results of Sec.~\ref{sec:results} carry over in closed form once such a channel is included, whereas the exact battery--controller correspondence of Sec.~\ref{sec:controller-cost} holds only in its absence. Since all results depend on the channel only through a single interarm overlap, the analysis applies to any path-dephasing channel of this form. We realize it here with the Wigner rotation of a massive particle's spin passing through the interferometers, following Ref.~\cite{pedro_basso_QCC_2024}.

Relativistic motion decoheres matter-wave interference through more than one channel~\cite{zych_ncomms_2011,pikovski2015}. When the particle carries an internal ``clock'', the differential proper time between the two arms dephases the clock states and, with them, the interference pattern~\cite{zych_ncomms_2011}. The channel we adopt is independent of that one and does not require a clock.
A massive particle with internal spin acquires along each path $\eta\in\{0,1\}$ a local Wigner rotation~\cite{pedro_basso_QCC_2024,Basso_2021,Terashima_PRA_2004}, $\ket{\tau_\eta}=\hat D[W(x_\eta,\tau_\eta)]\ket{\tau}$, sourced by the four-acceleration and the spin connection of that arm's world line. The effect is purely kinematic, appearing whether the two arms differ by relative motion in flat spacetime or by genuine curvature~\cite{pedro_basso_QCC_2024}. The spacetime geometry thereby entangles the path with the spin, and its entire effect on the path qubit is carried by the single complex overlap $\langle\tau_0|\tau_1\rangle$, which we combine with the interferometric phase into
\begin{equation}
    \Omega \equiv e^{i\phi}\langle\tau_0|\tau_1\rangle ,
    \qquad
    \Omega_R = \mathrm{Re}\,\Omega,
    \quad
    \Omega_I = \mathrm{Im}\,\Omega,
    \label{eq:Omega-def}
\end{equation}
with modulus $\nu\equiv|\langle\tau_0|\tau_1\rangle|
=\sqrt{\Omega_R^2+\Omega_I^2}$.
The non-relativistic analysis of the previous sections is recovered as $\langle\tau_0|\tau_1\rangle\to1$, i.e.\ $\Omega_R\to\cos\phi$, $\Omega_I\to\sin\phi$, $\nu\to1$.
We take the overlap real and positive, $\langle\tau_0|\tau_1\rangle=\nu$, as for planar motion (Appendix~\ref{app:rel}.\ref{app:wigner}), since a complex phase only shifts $\phi$.
The Gibbs initialization and the circuits are unchanged. The arm-dependent Wigner rotations act between the beam splitters, so tracing out the spin replaces $e^{\pm i\phi}$ by $\langle\tau_{0,1}|\tau_{1,0}\rangle\,e^{\pm i\phi}$ in the off-diagonal of the reduced battery state. After the full circuit, the output states retain the diagonal structure of the non-relativistic case with $\cos\phi\to\Omega_R$, while their coherences acquire the modulation given by Eq.~\eqref{eq:Omega-def}. The explicit states and derivations are collected in Appendix~\ref{app:rel}. We take the Wigner-rotation formalism from Ref.~\cite{pedro_basso_QCC_2024}, where $\langle\tau_0|\tau_1\rangle$ is obtained from the local Lorentz transformations along the two arms, and summarize the main aspects below.
\subsection{Wave--particle quantifiers}
\begin{figure*}[t]
    \centering
    \includegraphics[width=1\linewidth]{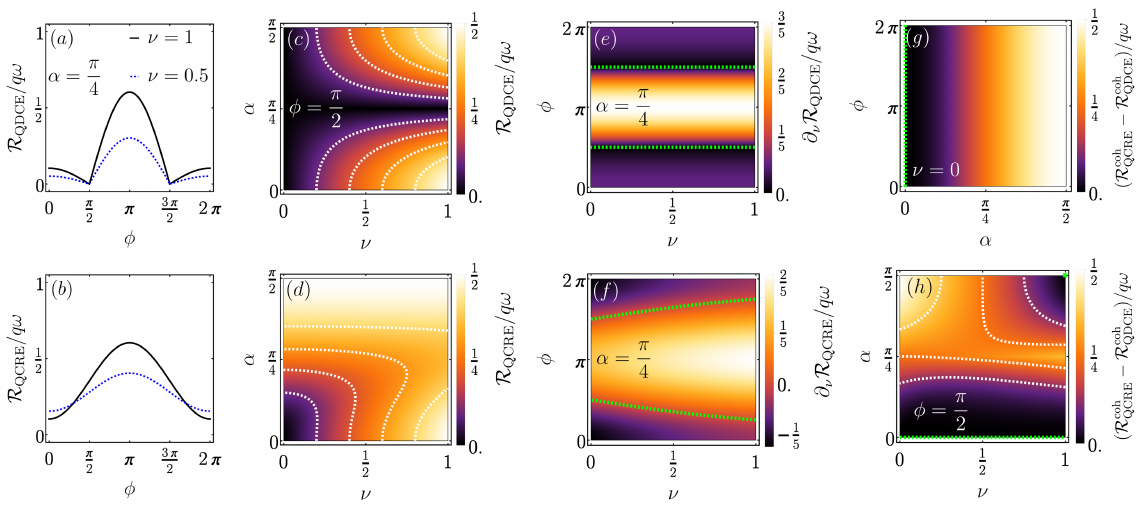}
\caption{Output ergotropy in the relativistic setting, normalized as $\Rcal/q\omega$ so that the thermal factor drops out and the curves hold at any temperature. The overlap is taken real and positive, $\langle\tau_0|\tau_1\rangle=\nu$. (a), (b) Versus the interferometric phase $\phi$ at $\alpha=\pi/4$, for the QDCE and the QCRE, in the non-relativistic limit $\nu=1$ (solid black) and at $\nu=|\langle\tau_0|\tau_1\rangle|=0.5$ (dashed blue). (c), (d) On the $(\nu,\alpha)$ plane at $\phi=\pi/2$. (e), (f) Response to the overlap, $\partial_\nu\Rcal$, on the $(\nu,\phi)$ plane at $\alpha=\pi/4$. (g) Coherent gap on the $(\alpha,\phi)$ plane at $\nu=0$. (h) Coherent gap on the $(\nu,\alpha)$ plane at $\phi=\pi/2$. In (c)--(h), dotted white curves are iso-contours and green dashed lines mark the zero loci.}
    \label{fig:rel-nonrel}
\end{figure*}
Of the three duality quantifiers, only the visibility is modified by the spacetime geometry. The predictability is fixed at the first beam splitter, before the Wigner rotations act, and therefore retains its non-relativistic value,
\begin{equation}
    P_{\mathrm{QDCE}} = 0,
    \qquad
    P_{\mathrm{QCRE}} = q\sin^2\!\alpha .
    \label{eq:P-rel}
\end{equation}
The visibility, shared by both setups, is suppressed by the Wigner-overlap modulus,
\begin{equation}
    \mathcal{V}=\nu q\cos^2\!\alpha\, = |\langle\tau_0|\tau_1\rangle|q\cos^2\!\alpha ,
    \label{eq:V-rel}
\end{equation}
so that the thermal factor $q$ and the geometric factor $\nu$ degrade the fringe contrast independently. The distinguishability is likewise untouched, since each path branch carries
a single spin label, so that the spin trace acts trivially in the path-projected operators and the overlap $\langle\tau_0|\tau_1\rangle$ does not enter $\mathcal{D}$. The QDCE
detector distinguishability vanishes as before, the particle content residing in the blocking channel, $\mathcal{D}^{\mathrm{ret}}_{\mathrm{QDCE}}=\sin^2\!\alpha$, while the QCRE retains its non-relativistic value
\begin{equation}
    \mathcal{D}_{\mathrm{QCRE}}
    = \sin\alpha\,\sqrt{q^2\sin^2\!\alpha+2\cos^2\!\alpha} .
    \label{eq:D-rel}
\end{equation}
Here $\mathcal D$ refers to the controller as the detector. The spin is itself a which-way marker: as $\nu\to0$ the spin states $\ket{\tau_0}$ and $\ket{\tau_1}$ become orthogonal, so the spin records the path perfectly in both protocols while $\mathcal V\to0$. The output ergotropy that the QCRE retains in this limit, Eq.~\eqref{eq:QCRE-remnant} below, therefore stems from the a priori predictability $P$, not from a which-way record.
Both the Englert and the D\"urr bounds keep the form of Sec.~\ref{sec:prelim}, with the visibility carrying the extra factor $\nu$. For the QDCE the predictability and the
detector distinguishability both vanish, so either combination reduces to $\mathcal{V}^2=q^2\nu^2\cos^4\!\alpha\le1$. For the QCRE,
\begin{align}
    \bigl(\mathcal{V}^2+\mathcal{D}^2\bigr)_{\mathrm{QCRE}}
    &= \frac{1+q^2}{2}-\frac{1-q^2}{2}\cos^2(2\alpha)\nonumber \\
      & - q^2(1-\nu^2)\cos^4\!\alpha\leq1,
    \label{eq:EGY-QCRE-rel}\\
    \bigl(P^2+\mathcal{V}^2\bigr)_{\mathrm{QCRE}}
    &= q^2\bigl(\sin^4\!\alpha+\nu^2\cos^4\!\alpha\bigr)\leq1,
    \label{eq:Durr-QCRE-rel}
\end{align}
The Englert combination being Eq.~\eqref{eq:EGY-QCRE} reduced by the geometric deficit $q^2(1-\nu^2)\cos^4\!\alpha$. The thermal and geometric contributions suppress saturation independently, since at zero temperature ($q\to1$) only the geometric deficit survives, while in the non-relativistic limit ($\nu\to1$) only the thermal one remains. The Englert bound is therefore saturated for any $\alpha$ once $q\to1$ and $\nu\to1$, whereas the D\"urr bound demands in addition $\alpha=\pi/2$, or $\alpha=0$ with $\nu\to1$.

\subsection{Ergotropy}
Introducing the relativistic generalization of $G(\alpha,\phi)$,
\begin{equation}
    G_{\mathrm{rel}}(\alpha;\Omega)
    \equiv \sqrt{(\cos^4\!\alpha+\sin^4\!\alpha)\,\Omega_R^2
        + \cos^2(2\alpha)\,\Omega_I^2},
    \label{eq:Grel-def}
\end{equation}
which reduces to $G(\alpha,\phi)$ as $\langle\tau_0|\tau_1\rangle\to1$, the ergotropic decomposition keeps the structure of Sec.~\ref{sec:results} with $\cos\phi$ replaced by~$\Omega_R$. The incoherent parts still coincide,
$\mathcal{R}^{\mathrm{inc}}_{\mathrm{QDCE}}=\mathcal{R}^{\mathrm{inc}}_{\mathrm{QCRE}}=\omega q\cos^2\!\alpha\,|\Omega_R|\,\Theta(-\Omega_R)$, while the coherent parts read
\begin{align}
    \Rcoh_{\mathrm{QDCE}}
      &= \frac{\omega q}{2}\bigl[G_{\mathrm{rel}}(\alpha;\Omega)
         -\cos^2\!\alpha\,|\Omega_R|\bigr],
    \label{eq:Rcoh-QDCE-rel}\\[2pt]
    \Rcoh_{\mathrm{QCRE}}
      &= \frac{\omega}{2}\bigl[\sqrt{\mathcal{V}^2+P^2}
         -q\cos^2\!\alpha\,|\Omega_R|\bigr].
    \label{eq:Rcoh-QCRE-rel}
\end{align}
The coherent gap, therefore, keeps the form of Eq.~\eqref{eq:Rcoh-diff},
\begin{equation}
    \Rcoh_{\mathrm{QCRE}}-\Rcoh_{\mathrm{QDCE}}
    = \frac{\omega}{2}\bigl[\sqrt{\mathcal{V}^2+P^2}
      -q\,G_{\mathrm{rel}}(\alpha;\Omega)\bigr]\ \ge 0 ,
    \label{eq:Rcoh-diff-rel}
\end{equation}
the inequality following, as in Sec.~\ref{sec:comparison}, from the sign of the difference of the squares,
\begin{equation}
    \mathcal{V}^2+P^2-q^2G_{\mathrm{rel}}^2
    = P^2\bigl(1-\nu^2\bigr)
      + \frac{q^2}{2}\sin^2(2\alpha)\,\Omega_I^2 \ \ge 0 ,
    \label{eq:gap-identity-rel}
\end{equation}
which collapses to Eq.~\eqref{eq:gap-identity} at $\nu\to1$, where $\Omega_I\to\sin\phi$. The coexistence term survives, now carried by $\Omega_I$, but a second, purely geometric term $\propto P^2(1-\nu^2)$ appears nonzero even where $\Omega_I=0$. The non-relativistic ordering thus survives the Wigner-induced decoherence, and the QCRE advantage now persists for $\alpha\neq0$ and $q>0$ on the lines $\phi\in\{0,\pi\}$ where the non-relativistic gap vanishes. Unlike the coexistence term, the geometric term carries no wave factor: at $\nu=0$, where $\mathcal{V}=0$, the gap reduces to $\tfrac{\omega}{2}P$. Under decoherence the QCRE advantage is therefore no longer a coexistence effect, and in the fully decohered limit it is a pure predictability effect.

The controller side does not carry over in the same way. The QCRE controller state is fixed at the first beam splitter and is unaffected by the spin, while the QDCE controller coherence acquires the factor $\Omega_R$, so that
\begin{equation}
    \psi^2_{C,\mathrm{QCRE}}-\psi^2_{C,\mathrm{QDCE}}
    =\frac{q^2}{2}\sin^2(2\alpha)\bigl(1-\Omega_R^2\bigr)\ \ge 0 .
    \label{eq:controller-purity-rel}
\end{equation}
The entropy ordering, Eq.~\eqref{eq:S-ordering}, therefore persists, but the right-hand side differs from Eq.~\eqref{eq:gap-identity-rel}: the geometric contribution is $\tfrac{q^2}{2}\sin^2(2\alpha)(1-\nu^2)$ rather than $P^2(1-\nu^2)$, and the battery--controller offset of Sec.~\ref{sec:controller-cost} becomes protocol dependent. The exact correspondence of Eq.~\eqref{eq:controller-purity-identityMAIN} is thus specific to $\nu=1$.

As mentioned previously, for $\langle\tau_0|\tau_1\rangle\to1$ all expressions reduce to the non-relativistic Gibbs results of Sec.~\ref{sec:results}. At $q\to1$ they recover the pure-input analysis of Ref.~\cite{pedro_basso_QCC_2024} with the overlap kept complex. At $q\to0$ every ergotropy vanishes.
The most revealing limit is that of orthogonal spins, $\nu\to0$, corresponding to a complete Wigner-induced decoherence. In this case, the QDCE ergotropy vanishes at both stages,
$\mathcal{R}^{\mathrm{in}}_{\mathrm{QDCE}} =\mathcal{R}_{\mathrm{QDCE}}\to0$, and so does the QCRE inside ergotropy,
\begin{equation}
    \mathcal{R}^{\mathrm{in}}_{\mathrm{QCRE}}\big|_{\nu\to0}
    = \tfrac{\omega}{2}\bigl(\sqrt{0+P^2}-P\bigr) = 0,
\end{equation}
yet the QCRE output ergotropy retains a finite remnant,
\begin{equation}
    \mathcal{R}_{\mathrm{QCRE}}\big|_{\nu\to0}
    = \frac{\omega q}{2}\sin^2\!\alpha
    = \frac{\omega}{2}\,P .
    \label{eq:QCRE-remnant}
\end{equation}
Complete decoherence therefore removes all extractable work inside the interferometer, while the fixed second beam splitter regenerates a finite output ergotropy because the $\ket{\mathrm{out}}$ branch of the QCRE carries a single spin label and survives the spin trace.  This output remnant separates the two protocols even on the lines $\phi\in\{0,\pi\}$, where their non-relativistic ergotropies coincide, so the Wigner decoherence breaks a degeneracy that the non-relativistic analysis cannot resolve.
Figure~\ref{fig:rel-nonrel} makes both effects visible. In panels (a) and (b) the QDCE ergotropy is suppressed by the overall factor $\nu$ at every phase,
$\Rcal_{\mathrm{QDCE}}=\nu\,\Rcal_{\mathrm{QDCE}}^{\nu=1}$, and still touches zero at $\phi=\pi/2$ and $3\pi/2$, whereas the QCRE responds non-monotonically, decreasing near
$\phi=\pi$ but increasing near $\phi=0$ and $2\pi$, so that the non-relativistic limit does not bound it from above. This increase is not a resource supplied by the motion. The dephasing conserves the battery energy between the beam splitters, but the fixed second beam splitter then deposits more energy, since the output population $p_1=\tfrac12(1-q\cos^2\!\alpha\,\Omega_R)$ grows as $\Omega_R$ decreases near $\phi=0$. Because both protocols share the same populations at every $\nu$, the comparison between them at fixed $\nu$ is unaffected. Its minimum stays above zero and rises towards the
remnant Eq.~\eqref{eq:QCRE-remnant}, equal to $\tfrac{\omega}{2}P=\frac{\omega q}{4}$ at this mixing angle, as $\nu\to0$. Panels (c) and (d) extend the comparison to the whole
$(\nu,\alpha)$ plane, where the QDCE is passive along the entire $\nu=0$ edge and along the line $\alpha=\pi/4$, while the QCRE vanishes only at the single corner $\nu=0$,
$\alpha=0$, at which $P$ vanishes as well.

Panels (e) and (f) make the asymmetry quantitative through the response $\partial_\nu\Rcal$. For the QDCE that response is independent of $\nu$ and non-negative, vanishing only on the lines $\phi=\pi/2$ and $3\pi/2$, so relativistic motion can only reduce the QDCE work, as the linear scaling noted above already implies. The QCRE response instead changes sign, being negative near $\phi=0$ and $2\pi$, where a smaller overlap yields more extractable work.
The two panels, therefore, separate the settings at which the motion must be suppressed from those at which it can be tolerated, or can even raise the output ergotropy at the price of a larger output energy. This information is not carried by the ergotropy maps themselves, which display the magnitude of the work rather than its sensitivity to the spacetime geometry.
Panels (g) and (h) isolate the gap on two slices. At $\nu=0$ the QDCE coherent sector is
extinguished and the gap collapses onto the predictability alone, $\Rcoh_{\mathrm{QCRE}}-\Rcoh_{\mathrm{QDCE}}=\tfrac{\omega}{2}P$, flat in $\phi$ and
closing only at $\alpha=0$, so complete loss of the interarm overlap removes the phase as a control parameter without removing the advantage. At $\phi=\pi/2$ the gap
$\tfrac{\omega q}{2}\bigl[\sqrt{\nu^2\cos^4\!\alpha+\sin^4\!\alpha}-\nu|\cos(2\alpha)|\bigr]$ closes along the $\alpha=0$ edge and at the isolated corner $\nu=1$, $\alpha=\pi/2$. The region of thermodynamic inequivalence thus does not shrink as $\nu$ decreases, which is what makes the QCRE the more robust charging scheme under relativistic motion.
\section{Conclusion}
\label{sec:conclusion}
We have analyzed a Mach--Zehnder nondegenerate path qubit that serves simultaneously as a quantum battery and as a wave--particle duality probe, comparing the quantum delayed-choice (QDCE) and quantum-controlled reality (QCRE) protocols for a thermal input. We showed that, although the two protocols share the same visibility and the same incoherent ergotropy at every point of the parameter space, their coherent ergotropies differ.
At the output, the QCRE never extracts less coherent work than the QDCE, and the closed-form margin between them is a coexistence term that requires both visibility and the additional path predictability that the QCRE alone carries, and is modulated by the interferometric phase. A priori which-way information is therefore not thermodynamically inert, and the ergotropic ordering of the two setups expresses their differing predictability rather than any difference in stored energy or fringe visibility. This battery-side ordering is fixed by the order in which the controlled mixing and the phase shift act, and survives if the controller is replaced by a classical random switch; the quantum coherence of the controller enters only through its reset cost.

We then turned to the beam-splitter controller, which acts as a which-way detector in the QCRE, and found that the same closed form reappears as a purity deficit acquired by the QDCE controller relative to the QCRE one. Whenever the controller is a reusable resource, this deficit gives the QDCE a Landauer bound on the local reset cost that is never lower than that of the QCRE. The two protocols also differ in when the controller acquires its entropy. The QCRE controller acquires its final entropy at the first beam splitter and keeps it unchanged thereafter, whereas the QDCE controller stays pure until the last step.  The energetic consequences of path information thus appear on both sides of the battery–controller system, as extractable work on the battery and as a reset cost on the controller.

Extending the analysis to a relativistic scenario, where a Wigner rotation of the particle's spin realizes a concrete, clock-independent decoherence channel that degrades path coherence during transit, we found that the QDCE ergotropy decreases monotonically as the spin overlap is reduced, so, for a real spin--path overlap, its non-relativistic value acts as an upper bound. The QCRE instead responds in a phase-dependent way, with relativistic dephasing suppressing its extractable work in some phase sectors and enhancing it, together with the output energy, in others. Since the channel degrades only the wave-like sector and leaves the predictability intact, the QDCE becomes passive in the complete-decoherence limit, while the QCRE retains a finite ergotropy set by its surviving path predictability and generated at the final beam splitter. The channel thereby separates the two protocols even on the parameter lines where their non-relativistic ergotropies coincide. In the fully decohered limit, where both visibilities vanish, the remaining advantage is a pure predictability effect rather than a coexistence one.

These results identify a priori which-way information as a thermodynamic resource, made accessible at the output by the coherent ergotropy of the battery, where the fringes do not reveal it, and accompanied on the detector side by a lower local reset cost of the controller. They also single out the QCRE as the more robust output battery, at equal visibility, under local reset, and under relativistic decoherence alike. Extending the framework to dynamical charging protocols, to multi-qubit batteries, and to a full accessible-information treatment of the controller as a which-way memory is a natural direction for future work.

\begin{acknowledgments}
T.A. acknowledges support from the National Science Centre, Poland (Maestro-21 contract no.2021/42/A/ST2/00356).
A.H.A.M. acknowledges support from the National Science Centre, Poland (SONATINA-9, grant agreement no. UMO-2025/56/C/ST2/00368 entitled “Modularne kwantowe urządzenia termiczne: integrowanie termicznych funkcjonalności”).
B.A. and L.P. acknowledge the support from IRA Programme (project no. FENG.02.01-IP.05-0006/23) financed by the FENG program 2021-2027, Priority FENG.02, Measure FENG.02.01., with the support of the FNP. P.R.D. acknowledges support from CNPq/MCTI/FNDTC 22/2024 (No. 446775/2024-0).
\end{acknowledgments}

\bibliography{bib}

\appendix
\section{Ergotropy for a two-level system: a brief review}
\label{app:ergotropy-review}
For completeness, we recall the derivation of Eqs.~\eqref{eq:Rinc}-\eqref{eq:Rcoh} starting from the spectral definition of ergotropy. Given a state $\hat \rho$ and a Hamiltonian $\hat{H}=\sum_k\epsilon_k\ket{\epsilon_k}\!\bra{\epsilon_k}$ with $\epsilon_1\le\epsilon_2\le\cdots\le\epsilon_d$, the ergotropy is
\begin{equation}
\mathcal{R}(\hat\rho,\hat{H})\;=\;\Tr(\hat\rho\hat{H})-\min_{\hat U}\Tr(\hat U \hat{\rho} \hat U^\dagger\hat{H}),
\end{equation}
where the minimum is taken over all unitaries and is achieved by the passive state $\hat \rho_p=\sum_k r_k\ket{\epsilon_k}\!\bra{\epsilon_k}$, obtained by pairing the largest eigenvalue $r_1\ge r_2\ge\cdots$ of $\hat \rho$ with the lowest energy $\epsilon_1$, and so on.
For a qubit with $\hat{H}=\epsilon_0\ket{0}\!\bra{0}+\epsilon_1\ket{1}\!\bra{1}$, $\epsilon_0\le\epsilon_1$, and described as $\hat{\rho}=(1-p)|0\rangle\langle0|+p|1\rangle\langle1|+\frac{\mathcal{C}}{2}\left(e^{i\varphi/2}|0\rangle\langle1|+h.c.\right)$, the eigenvalues of $\rho$ are
\begin{equation}
r_\pm\;=\;\tfrac{1}{2}\left(1\pm\Delta_\rho\right),\quad
\Delta_\rho\;=\;\sqrt{(1-2p)^2+\Cpath^{\,2}}.
\end{equation}
The mean energy is $\Tr(\hat \rho\hat{H})=(1-p)\epsilon_0+p\epsilon_1$, while the passive-state energy is $\Tr(\hat\rho_p\hat{H})=r_+\epsilon_0+r_-\epsilon_1$. By simply subtracting, one obtains
\begin{equation}
\mathcal{R}(\hat \rho,\hat{H})\;=\;(\epsilon_1-\epsilon_0)\left[\,p-\tfrac{1}{2}(1-\Delta_\rho)\right],
\label{eq:E-qubit-general}
\end{equation}
which vanishes iff $\hat \rho$ is passive, i.e.\ $\Cpath=0$ and $p\le 1/2$.
\section{Distinguishability of the QCRE}
\label{app:D-QCRE}

We derive Eq.~\eqref{eq:D-QCRE} from the trace-norm distinguishability Eq.~\eqref{eq:D-def}. Since the distinguishability asks how well the controller identifies the path, we work here with the joint battery--controller state rather than with the reduced battery state of the main text. Starting from the Gibbs input, Eq.~\eqref{eq:gibbs}, after the controlled first beam splitter and the phase shift this joint state is the convex mixture
\begin{equation}
        \hat{\Sigma}_2^{\mathrm{QCRE}}
    = \frac{1+q}{2}\,
      \ket{\Phi_2^{+}}\!\bra{\Phi_2^{+}}
    + \frac{1-q}{2}\,
      \ket{\Phi_2^{-}}\!\bra{\Phi_2^{-}},
\end{equation}
whose battery marginal, $\hat\rho_2^{\mathrm{QCRE}}=\mathrm{Tr}_C(\hat\Sigma_2^{\mathrm{QCRE}})$, is the state quoted in Sec.~\ref{sec:qcre-prelim}. The components, inherited from the
$\ket{0}$ and $\ket{1}$ Gibbs weights, are
\begin{align}
    \ket{\Phi_2^{+}}
    &= \frac{\cos\alpha}{\sqrt{2}}
       \bigl(\ket{0}+e^{i\phi}\ket{1}\bigr)\ket{\mathrm{in}}
    +\sin\alpha\,\ket{0}\ket{\mathrm{out}},
    \\[4pt]
    \ket{\Phi_2^{-}}
    &= \frac{\cos\alpha}{\sqrt{2}}
       \bigl(\ket{0}-e^{i\phi}\ket{1}\bigr)\ket{\mathrm{in}}
    +e^{i\phi}\sin\alpha\,\ket{1}\ket{\mathrm{out}}.
\end{align}
Projecting onto the path basis $\{\ket{0},\ket{1}\}$ gives the unnormalized path-conditioned controller states
$\ket{\zeta_j^{\pm}}=\langle j|\Phi_2^{\pm}\rangle$,
\begin{align}
    \ket{\zeta_0^{+}}
    &= \frac{\cos\alpha}{\sqrt2}\ket{\mathrm{in}}
      +\sin\alpha\,\ket{\mathrm{out}},
    &
    \ket{\zeta_1^{+}}
    &= \frac{e^{i\phi}\cos\alpha}{\sqrt2}\ket{\mathrm{in}},
    \label{eq:zeta-plus-B}
\end{align}
\begin{align}
    \ket{\zeta_0^{-}}
    &= \frac{\cos\alpha}{\sqrt2}\ket{\mathrm{in}},
    \\
    \ket{\zeta_1^{-}}
    &= -\frac{e^{i\phi}\cos\alpha}{\sqrt2}\ket{\mathrm{in}}+e^{i\phi}\sin\alpha\,\ket{\mathrm{out}}.
    \label{eq:zeta-minus-B}
\end{align}
The squared norms give the component-conditional path probabilities,
\begin{align}
    \langle\zeta_0^{+}|\zeta_0^{+}\rangle
    &= \langle\zeta_1^{-}|\zeta_1^{-}\rangle
    = \tfrac{1}{2}(1+\sin^2\!\alpha),
    \\
    \langle\zeta_1^{+}|\zeta_1^{+}\rangle
    &= \langle\zeta_0^{-}|\zeta_0^{-}\rangle
    = \tfrac{1}{2}\cos^2\!\alpha,
\end{align}
and thermal averaging, $w_j=\mathrm{Tr}((\ket{j}\!\bra{j}\otimes\hat{\mathbb{I}}_C) \hat\Sigma_2^{\mathrm{QCRE}})$, reproduce the path weights of Eq.~\eqref{eq:w-QCRE}. The unnormalized controller operators conditioned on each path are
\begin{equation}
    w_j\,\hat \rho_C^{(j)}
    = \frac{1+q}{2}\,\ket{\zeta_j^{+}}\!\bra{\zeta_j^{+}}
    + \frac{1-q}{2}\,\ket{\zeta_j^{-}}\!\bra{\zeta_j^{-}},
\end{equation}
which satisfy $\mathrm{Tr}(w_j\hat \rho_C^{(j)})=w_j$.  In the basis $\{\ket{\mathrm{in}},\ket{\mathrm{out}}\}$ their difference is
\begin{equation}
    A = w_0\,\hat \rho_C^{(0)}-w_1\,\hat \rho_C^{(1)}
    = \begin{pmatrix}
        0 & \dfrac{\sin\alpha\cos\alpha}{\sqrt2} \\[8pt]
        \dfrac{\sin\alpha\cos\alpha}{\sqrt2} & q\sin^2\!\alpha
      \end{pmatrix}.
    \label{eq:app-A-QCRE}
\end{equation}
Since $\mathrm{Tr}(A)=q\sin^2\!\alpha$ and $\det A=-\tfrac{1}{2}\sin^2\!\alpha\cos^2\!\alpha\le0$, the two eigenvalues have opposite signs, so the operator modulus gives
\begin{align}
\mathcal{D}_{\mathrm{QCRE}}
    = \mathrm{Tr}|A|
    =& \sqrt{[\mathrm{Tr}(A)]^2-4\det A}\\
    =& \sin\alpha\,\sqrt{q^2\sin^2\!\alpha+2\cos^2\!\alpha},
\end{align}
which is Eq.~\eqref{eq:D-QCRE}. For comparison, applying the symmetric distinguishability $\sqrt{1-|\langle\zeta_0^{+}|\zeta_1^{+}\rangle|^2}$ to the normalized states would give $\sqrt{2\sin^2\!\alpha/(1+\sin^2\!\alpha)}$, which differs from the correct value precisely because it discards the path-weight imbalance
$w_0\neq w_1$.
\section{Controller states and the purity identity }
\label{app:controller}
We derive Eqs.~\eqref{eq:rho-QDCE} and \eqref{eq:rho-QCRE} for the controller states and the purity identity given by Eq.~\eqref{eq:controller-purity-identityMAIN} of Sec.~\ref{sec:controller-cost}. Both protocols apply the same entangling gate and differ only in the battery state it acts on, so we perform the partial trace once for an arbitrary input and specialize afterwards.

Let $\hat\rho$ be any single-qubit battery state and let
$\hat{\mathcal{U}}_{cH}=\sum_{i}\hat X_i\otimes\hat P_i$, with
$\hat X_{\mathrm{in}}=\hat U_H$, $\hat X_{\mathrm{out}}=\hat{\mathbb{I}}$ and
$\hat P_i=\ket{i}\!\bra{i}$, act on $\hat\rho\otimes\ket{\mathrm{BS}}\!\bra{\mathrm{BS}}$.
Then
\begin{align}
    &\hat{\mathcal{U}}_{cH}\bigl(\hat \rho\otimes\ket{\mathrm{BS}}\!\bra{\mathrm{BS}}\bigr)\hat{\mathcal{U}}_{cH}^\dagger
    \nonumber\\
    &\quad= \sum_{i,j} \bigl(\hat X_i\hat \rho \hat X_j^\dagger\bigr)\otimes\bigl(\hat P_i\ket{\mathrm{BS}}\!\bra{\mathrm{BS}}\hat P_j\bigr).
\end{align}
Taking the partial trace over the battery replaces each factor $\hat X_i\hat\rho\hat X_j^\dagger$ above by its trace. The diagonal terms give $\mathrm{Tr}(\hat\rho)=1$ and $\mathrm{Tr}(\hat U_H\hat\rho\hat U_H)=1$, the latter because $\hat U_H^2=\hat{\mathbb{I}}$, while the off-diagonal terms give $t\equiv\mathrm{Tr}(\hat U_H\hat\rho)$, which is real because $\hat U_H$ and $\hat\rho$ are both Hermitian. Combining these with
$\hat P_{\mathrm{in}}\ket{\mathrm{BS}}\!\bra{\mathrm{BS}}\hat P_{\mathrm{in}}
=\cos^2\!\alpha\ket{\mathrm{in}}\!\bra{\mathrm{in}}$,
$\hat P_{\mathrm{out}}\ket{\mathrm{BS}}\!\bra{\mathrm{BS}}\hat P_{\mathrm{out}}
=\sin^2\!\alpha\ket{\mathrm{out}}\!\bra{\mathrm{out}}$ and
$\hat P_{\mathrm{in}}\ket{\mathrm{BS}}\!\bra{\mathrm{BS}}\hat P_{\mathrm{out}}
=\sin\alpha\cos\alpha\ket{\mathrm{in}}\!\bra{\mathrm{out}}$, the controller marginal
$\hat\rho_C:=\mathrm{Tr}_B\bigl(\hat{\mathcal{U}}_{cH}(\hat\rho\otimes
\ket{\mathrm{BS}}\!\bra{\mathrm{BS}})\hat{\mathcal{U}}_{cH}^\dagger\bigr)$ reads, in the
basis $\{\ket{\mathrm{in}},\ket{\mathrm{out}}\}$,
\begin{equation}
    \hat \rho_C
    = \begin{pmatrix}
    \cos^2\!\alpha & t\sin\alpha\cos\alpha \\[4pt]
    t\sin\alpha\cos\alpha & \sin^2\!\alpha
    \end{pmatrix}.
    \label{eq:controller-general}
\end{equation}
The diagonal is fixed by the beam-splitter setting alone, and the entire dependence on the battery state acted on by the entangling gate enters through the single real number $t=\tfrac{1}{\sqrt2}\bigl[(\rho_{00}-\rho_{11})+(\rho_{01}+\rho_{10})\bigr]$.
The entangling gate acts first, on the Gibbs input $\hat\rho=\hat\theta_\beta$ of Eq.~\eqref{eq:gibbs}, which is diagonal in the energy eigenbasis, so only the population difference contributes and $t=q/\sqrt2$. The phase shift and the final beam splitter act on the battery alone and, as shown at the end of this appendix, leave the controller marginal untouched, so Eq.~\eqref{eq:controller-general} already gives the output controller state, Eq.~\eqref{eq:rho-QCRE}.

 The entangling gate acts last, on the phase-shifted state $\hat\rho=\hat\rho_2^{\mathrm{QDCE}}$, whose diagonal is flat, so only the coherences contribute and $t=q\cos\phi/\sqrt2$. Equation~\eqref{eq:controller-general} then gives the output controller state, Eq.~\eqref{eq:rho-QDCE}.
 From Eq.~\eqref{eq:controller-general},
$\psi_C^2=2\mathrm{Tr}(\hat\rho_C^2)-1=\cos^2(2\alpha)+\sin^2(2\alpha)\,t^2$. Subtracting
the QDCE case from the QCRE case,
\begin{align}
    \psi_{C,\mathrm{QCRE}}^2-\psi_{C,\mathrm{QDCE}}^2
    &= \tfrac{q^2}{2}\sin^2(2\alpha)\bigl(1-\cos^2\!\phi\bigr)
    \nonumber\\
    &= 2 \mathcal{V}P\sin^2\!\phi ,
    \label{eq:controller-purity-identity}
\end{align}
which is the same function of $\alpha$ and $\phi$ that appears in the battery gap,
Eq.~\eqref{eq:gap-identity}.
For any bipartite state $\hat\Sigma_{BC}$ and any local unitary $\hat U_B$ acting on the battery alone, tracing out the battery is unaffected by $\hat U_B$.
Hence $\hat\rho_C$ can change only at the step where the entangling gate acts, since every other step is a unitary on the battery alone. In the QCRE, this is Step~1, so $\hat \rho_C^{\mathrm{QCRE}}$ becomes Eq.~\eqref{eq:rho-QCRE} immediately and is preserved through Steps~2 and~3. In the QDCE, this happens on Step~3, so $\hat \rho_C^{\mathrm{QDCE}}=\ket{\mathrm{BS}}\!\bra{\mathrm{BS}}$ remains throughout Steps~1 and ~2 and only becomes Eq.~\eqref{eq:rho-QDCE} at the output. This is the argument behind the timing result of Sec.~\ref{sec:controller-cost}.
\section{Relativistic treatment with a Gibbs initial state}
\label{app:rel}

We collect here the explicit states, wave--particle quantifiers, and ergotropies underlying the relativistic section of the main text. The path qubit is coupled, through the spacetime geometry, to the particle's internal spin. Following Ref.~\cite{pedro_basso_QCC_2024}, each path $\eta\in\{0,1\}$ carries the Wigner-rotated spin state

\begin{equation}
    \ket{\tau_\eta} := \hat D[W(x_\eta,\tau_\eta)]\ket{\tau},
    \label{eq:tau-eta-def}
\end{equation}
where $\hat D$ is the spin representation of the Wigner rotation $W$ acquired at the wave-packet position $x_\eta$ and proper time $\tau_\eta$ along that arm
and $\langle\tau_0|\tau_1\rangle$ is kept as a single complex inner product characterizing the spin--path entanglement. It is convenient to combine it with the interferometric phase, such that
\begin{equation}
    \Omega \equiv e^{i\phi}\langle\tau_0|\tau_1\rangle,
    \quad
    \Omega_R = \mathrm{Re}\,\Omega,
    \quad
    \Omega_I = \mathrm{Im}\,\Omega,
    \label{eq:Omega-def-app}
\end{equation}
with modulus $\nu\equiv|\langle\tau_0|\tau_1\rangle|=|\Omega|=\sqrt{\Omega_R^2+\Omega_I^2}$. As expected, the non-relativistic regime is recovered as $\langle\tau_0|\tau_1\rangle\to1$, i.e. $\Omega_R\to\cos\phi$, $\Omega_I\to\sin\phi$, $\nu\to1$. The Gibbs initialization, Eq.~\eqref{eq:gibbs}, and the circuits are unchanged, and the arm-dependent Wigner
rotations act between the beam splitters, mapping
$\ket{0}\ket{\tau}\mapsto\ket{0}\ket{\tau_0}$ and
$\ket{1}\ket{\tau}\mapsto e^{i\phi}\ket{1}\ket{\tau_1}$ during propagation.
\subsection{Wigner rotations and the spin--path overlap}
\label{app:wigner}
The relativistic effects enter our reduced (battery) states only through the complex overlap $\langle\tau_0|\tau_1\rangle$ of the arm-resolved spin states.
The origin of this overlap is purely kinematical. Transporting the spin along a worldline requires a local Lorentz transformation relating the tetrad frames at successive points, and the accumulated transformation is a Wigner rotation fixed by the four-acceleration and the spin connection of the metric. The derivation is given in full in Ref.~\cite{pedro_basso_QCC_2024}. For motion confined to a spatial plane the rotation generator points along a single fixed axis, so each arm contributes one accumulated rotation angle and the overlap is real, set by the difference of the two. We take this real overlap positive, $\langle\tau_0|\tau_1\rangle=\nu$, since a negative value would only shift $\phi$ by $\pi$.

The modulus $\nu$ therefore interpolates between two limits. For $\nu\to1$ the two spin states coincide, the spin--path correlation disappears, and the non-relativistic analysis is recovered, whereas for $\nu\to0$ the spin states are orthogonal and the Wigner-induced decoherence is complete. Between them, the scale is set by the geometry. In the Newtonian limit near Earth, Ref.~\cite{pedro_basso_QCC_2024} estimates $1-\nu\sim10^{-14}$ even for kilometer-scale arms, so the correction is negligible in the laboratory while becoming appreciable in a genuinely relativistic setting, such as near-horizon or high-velocity propagation.
Since every wave--particle and ergotropy result below depends on the geometry only through $\nu$, we treat it throughout as a free parameter on $[0,1]$.
\subsection{Quantum delayed-choice experiment}

The unitary $\hat U_H\otimes\hat{\mathbb{I}}_C$ acts on the battery alone, leaving
a product $\hat \rho_1^{\mathrm{QDCE}}\otimes\ket{\mathrm{BS}}\!\bra{\mathrm{BS}}$ with reduced battery state
\begin{equation}
    \hat \rho_1^{\mathrm{QDCE}} = \frac{1}{2}
    \begin{pmatrix} 1 & q \\ q & 1 \end{pmatrix},
    \label{eq:rho1-QDCE-rel}
\end{equation}
identical to the non-relativistic case.
The phase shift imprints $\phi$ on arm~1 while the arm-dependent Wigner rotations entangle path with spin.  Tracing out the spin degree of freedom leads to
\begin{equation}
    \hat \rho_2^{\mathrm{QDCE}} = \frac{1}{2}
    \begin{pmatrix}
    1 & q\,e^{-i\phi}\langle\tau_1|\tau_0\rangle \\[2pt]
    q\,e^{i\phi}\langle\tau_0|\tau_1\rangle & 1
    \end{pmatrix},
    \label{eq:rho2-QDCE-rel}
\end{equation}
with the off-diagonal elements modulated by the complex Wigner overlap.
The joint unitary $\hat{\mathcal{U}}_{cH}$ produces
\begin{equation}
    \hat\Sigma_f^{\mathrm{QDCE}} =
    \tfrac{1+q}{2}\ket{\Psi_f^{+}}\!\bra{\Psi_f^{+}} +
    \tfrac{1-q}{2}\ket{\Psi_f^{-}}\!\bra{\Psi_f^{-}},
\end{equation}
with
\begin{align}
    \ket{\Psi_f^{+}} &= \tfrac{\cos\alpha}{\sqrt{2}}\bigl(\ket{x_+}\ket{\tau_0}
     + e^{i\phi}\ket{x_-}\ket{\tau_1}\bigr)\ket{\mathrm{in}} \nonumber\\
    &\quad + \tfrac{\sin\alpha}{\sqrt{2}}\bigl(\ket{0}\ket{\tau_0}
     + e^{i\phi}\ket{1}\ket{\tau_1}\bigr)\ket{\mathrm{out}},
    \\
    \ket{\Psi_f^{-}} &= \tfrac{\cos\alpha}{\sqrt{2}}\bigl(\ket{x_+}\ket{\tau_0}
     - e^{i\phi}\ket{x_-}\ket{\tau_1}\bigr)\ket{\mathrm{in}} \nonumber\\
    &\quad + \tfrac{\sin\alpha}{\sqrt{2}}\bigl(\ket{0}\ket{\tau_0}
     - e^{i\phi}\ket{1}\ket{\tau_1}\bigr)\ket{\mathrm{out}}.
\end{align}
Tracing out the spin and the controller degrees of freedom yields
\begin{equation}
    \hat \rho_S^{\mathrm{QDCE}} = \tfrac{1}{2}
    \begin{pmatrix} \rho_{00} & \rho_{01} \\ \rho_{01}^{\ast} & \rho_{11} \end{pmatrix},
    \label{eq:rhoS-QDCE-rel}
\end{equation}
with, using Eq.~\eqref{eq:Omega-def-app},
$\rho_{00} = 1 + q\cos^2\!\alpha\,\Omega_R$,
$\rho_{11} = 1 - q\cos^2\!\alpha\,\Omega_R$, and
$\rho_{01} = q(\sin^2\!\alpha\,\Omega_R + i\cos(2\alpha)\,\Omega_I)$.
\subsection{Quantum-controlled reality experiment}
The joint unitary $\hat{\mathcal{U}}_{cH}^{(1)}$ acts first, entangling battery and controller. Thus, the reduced battery state becomes
\begin{equation}
    \hat \rho_1^{\mathrm{QCRE}} = \tfrac{1}{2}
    \begin{pmatrix}
    1+q\sin^2\!\alpha & q\cos^2\!\alpha \\[2pt]
    q\cos^2\!\alpha & 1-q\sin^2\!\alpha
    \end{pmatrix}.
    \label{eq:rho1-QCRE-rel}
\end{equation}
After tracing out the spin degree of freedom,
\begin{equation}
    \hat \rho_2^{\mathrm{QCRE}} = \tfrac{1}{2}
    \begin{pmatrix}
    1+q\sin^2\!\alpha & q\,e^{-i\phi}\langle\tau_1|\tau_0\rangle\cos^2\!\alpha \\[2pt]
    q\,e^{i\phi}\langle\tau_0|\tau_1\rangle\cos^2\!\alpha & 1-q\sin^2\!\alpha
    \end{pmatrix},
    \label{eq:rho2-QCRE-rel}
\end{equation}
with the off-diagonal carrying the extra factor $\langle\tau_1|\tau_0\rangle$. The battery--controller system is correlated at this stage, encoding which-way information in the controller.
The final $\hat U_H\otimes\hat{\mathbb{I}}_C$ produces
\begin{equation}
    \hat\Sigma_f^{\mathrm{QCRE}} =
    \tfrac{1+q}{2}\ket{\Xi_f^{+}}\!\bra{\Xi_f^{+}} +
    \tfrac{1-q}{2}\ket{\Xi_f^{-}}\!\bra{\Xi_f^{-}},
\end{equation}
with
\begin{align}
    \ket{\Xi_f^{+}} &= \tfrac{\cos\alpha}{\sqrt{2}}\bigl(\ket{x_+}\ket{\tau_0}
     + e^{i\phi}\ket{x_-}\ket{\tau_1}\bigr)\ket{\mathrm{in}} \nonumber\\
    &\quad + \sin\alpha\,\ket{x_+}\ket{\tau_0}\ket{\mathrm{out}},
    \\
    \ket{\Xi_f^{-}} &= \tfrac{\cos\alpha}{\sqrt{2}}\bigl(\ket{x_+}\ket{\tau_0}
     - e^{i\phi}\ket{x_-}\ket{\tau_1}\bigr)\ket{\mathrm{in}} \nonumber\\
    &\quad + e^{i\phi}\sin\alpha\,\ket{x_-}\ket{\tau_1}\ket{\mathrm{out}}.
\end{align}
Tracing out the spin and the controller degrees of freedom gives
\begin{equation}
    \hat \rho_S^{\mathrm{QCRE}} = \tfrac{1}{2}
    \begin{pmatrix} \rho_{00} & \rho_{01} \\ \rho_{01}^{\ast} & \rho_{11} \end{pmatrix},
    \label{eq:rhoS-QCRE-rel}
\end{equation}
with, using Eq.~\eqref{eq:Omega-def-app},
$\rho_{00} = 1 + q\cos^2\!\alpha\,\Omega_R$,
$\rho_{11} = 1 - q\cos^2\!\alpha\,\Omega_R$, and
$\rho_{01} = q(\sin^2\!\alpha + i\cos^2\!\alpha\,\Omega_I)$.
The diagonal coincides with that of $\hat \rho_S^{\mathrm{QDCE}}$, i.e., the two setups share the same output population but differ in the off-diagonal
elements.
\subsection{Wave--particle quantifiers}
Of the three duality quantifiers, only the visibility is modified by the spacetime geometry. The predictability is fixed at the first beam splitter, before the Wigner rotations act, and is read from the populations of $\hat\rho_1^{\mathrm{QDCE}}$ and $\hat\rho_1^{\mathrm{QCRE}}$, Eqs.~\eqref{eq:rho1-QDCE-rel} and~\eqref{eq:rho1-QCRE-rel}. The subsequent rotations, $\ket{0}\ket{\tau}\mapsto\ket{0}\ket{\tau_0}$ and $\ket{1}\ket{\tau}\mapsto e^{i\phi}\ket{1}\ket{\tau_1}$, attach a single spin label to each path, so that tracing out the spin leaves the diagonal elements of $\hat\rho_2^{\mathrm{QDCE}}$ and $\hat\rho_2^{\mathrm{QCRE}}$ unchanged and only the off-diagonal elements acquire the overlap $\langle\tau_0|\tau_1\rangle$, Eqs.~\eqref{eq:rho2-QDCE-rel} and~\eqref{eq:rho2-QCRE-rel}. The predictability therefore retains its non-relativistic value,
\begin{equation}
    P_{\mathrm{QDCE}} = 0,\qquad
    P_{\mathrm{QCRE}} = q\sin^2\!\alpha.
    \label{eq:P-rel-app}
\end{equation}

The output port-0 probability $p_0 = \tfrac{1}{2}(1 + q\cos^2\!\alpha\,\Omega_R)$ coincides for both setups. Since the phase of $\langle\tau_0|\tau_1\rangle$ only shifts the fringes, extremizing over $\phi$ gives
\begin{equation}
    \mathcal{V}_{\mathrm{QDCE}} = \mathcal{V}_{\mathrm{QCRE}}
    = \nu q\cos^2\!\alpha,
    \label{eq:V-rel-app}
\end{equation}
so that the thermal factor $q$ and the Wigner-overlap modulus $\nu$ degrade the fringe contrast independently.

The controller remains in $\ket{\mathrm{BS}}$ throughout Steps~1 and~2, uncorrelated with both path and spin, so it carries no path information and the detector distinguishability vanishes, $\mathcal{D}_{\mathrm{QDCE}}=0$. The particle content is then probed by the blocking protocol, in which one arm is blocked after the Wigner rotations, the controlled second beam splitter is kept active, and the two output probabilities are compared. Blocking projects onto a definite path carrying a single spin label, so the surviving state is a pure path eigenstate and the overlap $\langle\tau_0|\tau_1\rangle$ does not enter. Propagating through $\hat{\mathcal{U}}_{cH}$ and averaging over the controller gives $\sin^2\!\alpha$ for either blocked arm, hence
\begin{equation}
    \mathcal{D}^{\mathrm{ret}}_{\mathrm{QDCE}} = \sin^2\!\alpha ,
    \label{eq:D-QDCE-block-rel}
\end{equation}
which is independent of $q$, $\phi$, and $\nu$.

Here, the controller is genuinely correlated with the path and acts as a which-way detector. The which-way information is read from the intermediate joint state, the convex mixture
\begin{equation}
    \hat\Sigma_2^{\mathrm{QCRE}}
    = \tfrac{1+q}{2}\ket{\Phi_2^{+}}\!\bra{\Phi_2^{+}}
    + \tfrac{1-q}{2}\ket{\Phi_2^{-}}\!\bra{\Phi_2^{-}},
\end{equation}
whose components, with the spin attached to each path, are
\begin{align}
    \ket{\Phi_2^{+}} &= \tfrac{\cos\alpha}{\sqrt2}
       \bigl(\ket{0}\ket{\tau_0}+e^{i\phi}\ket{1}\ket{\tau_1}\bigr)\ket{\mathrm{in}}
    \nonumber\\
    &\quad + \sin\alpha\,\ket{0}\ket{\tau_0}\ket{\mathrm{out}},
    \\
    \ket{\Phi_2^{-}} &= \tfrac{\cos\alpha}{\sqrt2}
       \bigl(\ket{0}\ket{\tau_0}-e^{i\phi}\ket{1}\ket{\tau_1}\bigr)\ket{\mathrm{in}}
    \nonumber\\
    &\quad + e^{i\phi}\sin\alpha\,\ket{1}\ket{\tau_1}\ket{\mathrm{out}}.
\end{align}
Projecting onto a definite path determines whether the overlap survives. Applying $\bra{0}$ to the battery slot of each component gives
\begin{align}
    \braket{0|\Phi_2^{+}}
    &= \ket{\tau_0}\otimes\Bigl(\tfrac{\cos\alpha}{\sqrt2}\ket{\mathrm{in}}
       + \sin\alpha\,\ket{\mathrm{out}}\Bigr),
    \\
    \braket{0|\Phi_2^{-}}
    &= \ket{\tau_0}\otimes\tfrac{\cos\alpha}{\sqrt2}\ket{\mathrm{in}},
\end{align}
and applying $\bra{1}$ gives
\begin{align}
    \braket{1|\Phi_2^{+}}
    &= \ket{\tau_1}\otimes\tfrac{e^{i\phi}\cos\alpha}{\sqrt2}\ket{\mathrm{in}},
    \\
    \braket{1|\Phi_2^{-}}
    &= \ket{\tau_1}\otimes\Bigl(-\tfrac{e^{i\phi}\cos\alpha}{\sqrt2}\ket{\mathrm{in}}
       + e^{i\phi}\sin\alpha\,\ket{\mathrm{out}}\Bigr).
\end{align}
Each path-projected piece factorizes into a single spin label times a controller state, with path $\ket{0}$ carrying $\ket{\tau_0}$ and path $\ket{1}$ carrying $\ket{\tau_1}$. Since the spin is perfectly correlated with the path in $\hat\Sigma_2^{\mathrm{QCRE}}$, the two spin labels are never superposed within a fixed path projection, and the spin trace contributes only $\langle\tau_0|\tau_0\rangle=\langle\tau_1|\tau_1\rangle=1$. The cross overlap $\langle\tau_0|\tau_1\rangle$ therefore never appears, and the path-conditioned controller states reduce to their non-relativistic form, Eqs.~\eqref{eq:zeta-plus-B}--\eqref{eq:zeta-minus-B},
\begin{align}
    \ket{\zeta_0^{+}} &= \tfrac{\cos\alpha}{\sqrt{2}}\ket{\mathrm{in}}
    + \sin\alpha\,\ket{\mathrm{out}},
    &
    \ket{\zeta_1^{+}} &= \tfrac{e^{i\phi}\cos\alpha}{\sqrt{2}}\ket{\mathrm{in}},
\end{align}
\begin{align}
    \ket{\zeta_0^{-}} &= \tfrac{\cos\alpha}{\sqrt{2}}\ket{\mathrm{in}},
    &
    \ket{\zeta_1^{-}} &= -\tfrac{e^{i\phi}\cos\alpha}{\sqrt{2}}\ket{\mathrm{in}}
     + e^{i\phi}\sin\alpha\,\ket{\mathrm{out}}.
    \label{eq:zeta-QCRE-rel-app}
\end{align}

The weighted conditional controller operators are
\begin{align}
    w_j\,\hat \rho_C^{(j)} = \tfrac{1+q}{2}\ket{\zeta_j^{+}}\!\bra{\zeta_j^{+}}
     \,+ \tfrac{1-q}{2}\ket{\zeta_j^{-}}\!\bra{\zeta_j^{-}},
\end{align}
with squared norms reproducing the path weights of Eq.~\eqref{eq:w-QCRE}, so that $w_0-w_1=q\sin^2\!\alpha=P_{\mathrm{QCRE}}$. Applying the asymmetric trace-norm distinguishability, Eq.~\eqref{eq:D-def}~\cite{englert_prl_1996}, and regrouping the difference by Gibbs component,
\begin{equation}
    A := w_0\hat \rho_C^{(0)} - w_1\hat \rho_C^{(1)}
    = \tfrac{1+q}{2}\Delta^{+} + \tfrac{1-q}{2}\Delta^{-},
\end{equation}
with $\Delta^{\pm}:=\ket{\zeta_0^{\pm}}\!\bra{\zeta_0^{\pm}}-\ket{\zeta_1^{\pm}}\!\bra{\zeta_1^{\pm}}$, direct evaluation in the $\{\ket{\mathrm{in}},\ket{\mathrm{out}}\}$ basis gives
\begin{equation}
    \Delta^{\pm} = \begin{pmatrix} 0 & \sin\alpha\cos\alpha/\sqrt{2} \\[3pt]
    \sin\alpha\cos\alpha/\sqrt{2} & \pm\sin^2\!\alpha \end{pmatrix}.
\end{equation}
The off-diagonal elements of $\Delta^{+}$ and $\Delta^{-}$ coincide while their $\ket{\mathrm{out}}\!\bra{\mathrm{out}}$ entries are opposite, so only the thermal residue survives there, and $A$ reproduces Eq.~\eqref{eq:app-A-QCRE},
\begin{equation}
    A = \begin{pmatrix}
    0 & \sin\alpha\cos\alpha/\sqrt{2}\\[4pt]
    \sin\alpha\cos\alpha/\sqrt{2} & q\sin^2\!\alpha
    \end{pmatrix}.
    \label{eq:A-rel-app}
\end{equation}
Since $\mathrm{Tr}(A)=q\sin^2\!\alpha$ and $\det A=-\tfrac12\sin^2\!\alpha\cos^2\!\alpha\le0$, the eigenvalues have opposite signs, and
\begin{align}
    \mathcal{D}_{\mathrm{QCRE}} &= \mathrm{Tr}|A| = \sqrt{[\mathrm{Tr}(A)]^2-4\det A}
    \nonumber\\
    &= \sin\alpha\,\sqrt{q^2\sin^2\!\alpha + 2\cos^2\!\alpha},
    \label{eq:D-QCRE-rel-app}
\end{align}
unchanged from the non-relativistic value, Eq.~\eqref{eq:D-QCRE}. The overlap $\nu$ thus enters only the interference sector, namely the visibility and the coherence, and never the path-conditioned distinguishability, since only interference superposes the two spin labels. Here $\mathcal{D}$ refers to the controller as the accessible which-way detector, with the spin traced out as an inaccessible internal record. A joint controller--spin readout would acquire a dependence on $\nu$.

For the QDCE the predictability and the detector distinguishability vanish, and the particle content enters through the blocking value, Eq.~\eqref{eq:D-QDCE-block-rel}, giving the retrodictive relation
\begin{equation}
    \mathcal{V}^2+\bigl(\mathcal{D}^{\mathrm{ret}}_{\mathrm{QDCE}}\bigr)^2
    = q^2\nu^2\cos^4\!\alpha + \sin^4\!\alpha \le 1,
    \label{eq:ret-QDCE-rel-app}
\end{equation}
the relativistic counterpart of Eq.~\eqref{eq:EGY-ret}, with the visibility carrying the extra factor $\nu$. For the QCRE, substituting Eqs.~\eqref{eq:P-rel-app}, \eqref{eq:V-rel-app} and~\eqref{eq:D-QCRE-rel-app} into the Englert and D\"urr combinations gives Eqs.~\eqref{eq:EGY-QCRE-rel} and~\eqref{eq:Durr-QCRE-rel} of the main text. The deficit from saturation of the Englert bound splits into a thermal and a geometric contribution,
\begin{align}
    1-\bigl(\mathcal{V}^2+\mathcal{D}^2\bigr)_{\mathrm{QCRE}}
    &= \tfrac{1-q^2}{2}\bigl(1+\cos^2(2\alpha)\bigr)
    \nonumber\\
    &\quad + q^2(1-\nu^2)\cos^4\!\alpha,
    \label{eq:EGY-gap-rel-app}
\end{align}
where the first term arises from the thermal mixedness of the input and the second from the Wigner-induced decoherence. Both terms vanish for $q\to1$ and $\nu\to1$, so the Englert bound is saturated for any $\alpha$ in that limit, whereas the D\"urr combination, Eq.~\eqref{eq:Durr-QCRE-rel}, reaches unity only for $q\to1$ at $\alpha=\pi/2$, or at $\alpha=0$ with $\nu\to1$.
\subsection{Ergotropy}
We use the relativistic generalization $G_{\mathrm{rel}}(\alpha;\Omega)$ of $G(\alpha,\phi)$ introduced in Eq.~\eqref{eq:Grel-def}, which reduces to $G(\alpha,\phi)$ as $\langle\tau_0|\tau_1\rangle\to1$.
\subsubsection*{QDCE}
Inside the interferometer, Eq.~\eqref{eq:rho2-QDCE-rel} gives $p_{1,\mathrm{in}}=\tfrac12$ and $\mathcal{C}_{\mathrm{in}}=\psi_{\mathrm{in}}=q\nu$, so $\Rinc_{\mathrm{in}}=0$ and the ergotropy is purely coherent,
\begin{equation}
    \mathcal{R}^{\mathrm{in}}_{\mathrm{QDCE}}
    = \tfrac{\omega q}{2}\,\nu.
\end{equation}
At the output, Eq.~\eqref{eq:rhoS-QDCE-rel} gives
\begin{align}
    p_1 &= \tfrac{1}{2}(1 - q\cos^2\!\alpha\,\Omega_R),\\
    \mathcal{C} &= q\sqrt{\sin^4\!\alpha\,\Omega_R^2 + \cos^2(2\alpha)\,\Omega_I^2},\\
    \psi &= q\,G_{\mathrm{rel}}(\alpha;\Omega),
\end{align}
and the identity $\psi^2-\mathcal{C}^2=q^2\cos^4\!\alpha\,\Omega_R^2$ gives $\sqrt{\psi^2-\mathcal{C}^2}=q\cos^2\!\alpha\,|\Omega_R|$. Since $p_1\ge\tfrac12$ if and only if $\Omega_R\le0$,
\begin{align}
    \Rinc_{\mathrm{QDCE}}
    &= \omega q\cos^2\!\alpha\,|\Omega_R|\,\Theta(-\Omega_R),\\
    \Rcoh_{\mathrm{QDCE}}
    &= \tfrac{\omega q}{2}\bigl[G_{\mathrm{rel}}(\alpha;\Omega)
       - \cos^2\!\alpha\,|\Omega_R|\bigr],
\end{align}
which combine into the closed-form total, valid for all $\phi$,
\begin{equation}
    \mathcal{R}_{\mathrm{QDCE}}
    = \tfrac{\omega q}{2}\bigl[G_{\mathrm{rel}}(\alpha;\Omega)
       - \cos^2\!\alpha\,\Omega_R\bigr].
    \label{eq:R-total-QDCE-rel-app}
\end{equation}
Since $\Omega_R$ and $\Omega_I$ are both proportional to $\nu$, this total is linear in the overlap, $\mathcal{R}_{\mathrm{QDCE}}=\nu\,\mathcal{R}_{\mathrm{QDCE}}^{\nu=1}$ at fixed $\phi+\arg\langle\tau_0|\tau_1\rangle$.
\subsubsection*{QCRE}
Inside the interferometer, Eq.~\eqref{eq:rho2-QCRE-rel} gives $p_{1,\mathrm{in}}=\tfrac12(1-P)$, $\mathcal{C}_{\mathrm{in}}=\nu q\cos^2\!\alpha=\mathcal{V}$, and $\psi_{\mathrm{in}}=q\sqrt{\sin^4\!\alpha+\nu^2\cos^4\!\alpha}=\sqrt{\mathcal{V}^2+P^2}$. The identity $\psi_{\mathrm{in}}^2-\mathcal{C}_{\mathrm{in}}^2=P^2$ is preserved, since both terms acquire the same factor $\nu^2$ in their $\cos^4\!\alpha$ part, so $\Rinc_{\mathrm{in}}=0$ and
\begin{equation}
    \mathcal{R}^{\mathrm{in}}_{\mathrm{QCRE}}
    = \tfrac{\omega}{2}\bigl(\sqrt{\mathcal{V}^2+P^2}-P\bigr).
\end{equation}
At the output, Eq.~\eqref{eq:rhoS-QCRE-rel} gives
\begin{align}
    p_1 &= \tfrac{1}{2}(1 - q\cos^2\!\alpha\,\Omega_R),\\
    \mathcal{C} &= q\sqrt{\sin^4\!\alpha + \cos^4\!\alpha\,\Omega_I^2},\\
    \psi &= \sqrt{\mathcal{V}^2+P^2},
\end{align}
with $p_1$ equal to that of the QDCE and $\psi$ independent of $\phi$. The identity $\psi^2-\mathcal{C}^2=q^2\cos^4\!\alpha\,\Omega_R^2$ matches the QDCE output, so the incoherent ergotropies coincide, $\Rinc_{\mathrm{QCRE}}=\Rinc_{\mathrm{QDCE}}=\omega q\cos^2\!\alpha\,|\Omega_R|\,\Theta(-\Omega_R)$, while
\begin{equation}
    \mathcal{R}_{\mathrm{QCRE}}
    = \tfrac{\omega}{2}\bigl[\sqrt{\mathcal{V}^2+P^2}
       - q\cos^2\!\alpha\,\Omega_R\bigr],
    \label{eq:R-total-QCRE-rel-app}
\end{equation}
valid for all $\phi$.

\subsection{Limits and special cases}

\paragraph*{Non-relativistic limit, $\langle\tau_0|\tau_1\rangle\to1$.}
$\Omega_R\to\cos\phi$, $\Omega_I\to\sin\phi$, $\nu\to1$, $G_{\mathrm{rel}}\to G(\alpha,\phi)$, and all quantities reduce to the Gibbs results of the main text.

\paragraph*{Pure input, $q\to1$.}
The thermal prefactor saturates and the results reduce to their pure-input counterparts, recovering Ref.~\cite{pedro_basso_QCC_2024} with the overlap kept complex.

\paragraph*{Complete dephasing, $\nu\to0$.}
$\Omega_R,\Omega_I\to0$, $G_{\mathrm{rel}}\to0$, and $\mathcal{V}\to0$. The QDCE ergotropy vanishes at both stages, $\mathcal{R}^{\mathrm{in}}_{\mathrm{QDCE}}=\mathcal{R}_{\mathrm{QDCE}}\to0$. For the QCRE the inside ergotropy also vanishes, $\mathcal{R}^{\mathrm{in}}_{\mathrm{QCRE}}=\tfrac{\omega}{2}\bigl(\sqrt{0+P^2}-P\bigr)=0$, while the output retains a finite remnant set by the predictability,
\begin{equation}
    \mathcal{R}_{\mathrm{QCRE}}\big|_{\nu\to0}
    = \tfrac{\omega q}{2}\sin^2\!\alpha = \tfrac{\omega}{2}\,P,
    \label{eq:QCRE-remnant-app}
\end{equation}
because the $\ket{\mathrm{out}}$ branches of $\ket{\Xi_f^{\pm}}$ each carry a single spin label that survives the spin trace, and the fixed second beam splitter converts the surviving amplitude into a nonzero output coherence.

\paragraph*{Maximally mixed input, $q\to0$.}
All four ergotropies vanish, as expected for the maximally mixed input.

\paragraph*{Closed interferometer, $\alpha=0$.}
Both setups reduce to a standard MZI with $P=\mathcal{D}=0$ and $\mathcal{V}=\nu q$, and their ergotropies coincide,
\begin{equation}
    \mathcal{R}\big|_{\alpha=0}
    = \tfrac{\omega q}{2}\bigl(\nu - \Omega_R\bigr).
\end{equation}

\paragraph*{Open interferometer, $\alpha=\pi/2$.}
The controller is always in $\ket{\mathrm{out}}$ and no controlled beam splitter acts. The QDCE output then keeps the coherence created by the fixed first beam splitter, suppressed by the overlap, $\mathcal{R}_{\mathrm{QDCE}}|_{\alpha=\pi/2}=\tfrac{\omega q}{2}\,\nu$, whereas the QCRE output is insensitive to the geometry, $\mathcal{R}_{\mathrm{QCRE}}|_{\alpha=\pi/2}=\tfrac{\omega q}{2}$. The two setups, which coincide at $\omega q/2$ in the non-relativistic limit, are therefore separated by $\tfrac{\omega q}{2}(1-\nu)$.

\end{document}